\documentclass[aps,pra,showpacs,twoside,10pt,floatfix,nofootinbib,longbibliography,twocolumn]{revtex4-2}
\usepackage[colorlinks=true, citecolor=red, urlcolor=blue ]{hyperref}
\usepackage{times,epsfig,amssymb,amsfonts,amsmath,bm,subfigure,mathtools,amsthm,braket,soul,enumitem,color,physics,graphics,graphicx}
\usepackage[normalem]{ulem}
\usepackage{comment}
\usepackage{epstopdf}

\newcommand{\adi}[1]{{\color{black}#1}}
\usepackage{optidef}

\begin{document}

\title{Hierarchies among genuine multipartite entangling capabilities of quantum gates}

\author{ Mrinmoy Samanta\(^1\), Sudipta Mondal\(^1\), Samir Kumar Hazra\(^{1,2}\), Aditi Sen(De)\(^1\)}
 \affiliation{\(^1\)Harish-Chandra Research Institute, A CI of Homi Bhabha National Institute,  Chhatnag Road, Jhunsi, Allahabad - 211019, India\\
\(^2\) MURTI - Quantum Information Lab and Department of Mathematics, Gandhi Institute of Technology and Management (Deemed to be University), Bengaluru-562163, Karnataka, India
}

\begin{abstract}

We classify quantum gates according to their capability to generate genuine multipartite entanglement (GME), using a hierarchy based on multipartite separable states. 
 In particular, when a fixed unitary operator acts on the set of  \(k\)-separable states, the maximal (average) genuine multipartite entanglement content produced via that particular unitary operator is determined after maximizing over the set of \(k\)-separable input states. We identify unitary operators that are beneficial for generating high GME when the input states are entangled in some bipartition, although the picture can also be reversed where such initial entanglement offers no advantage. We investigate the maximum entangling power of a broad range of unitary operators, encompassing special classes of quantum gates, as well as diagonal, permutation, and Haar-uniformly generated unitaries by computing the generalized geometric measure  (GGM) as GME quantifier. \adi{Additionally, we observe a notable distinction in entangling power based on the nature of the input states: when maximization is restricted to separable states with real coefficients, the entangling power is lower than when the optimization is carried out over arbitrary separable states with complex coefficients, thereby highlighting the role of complex amplitudes in entanglement creation.} Furthermore, we determine which unitary operators, along with their corresponding optimal inputs, yield output states with the highest achievable GGM.  
  

\end{abstract}
\maketitle

\section{Introduction}
\label{sec:intro}


Quantum correlations \cite{Horodecki2009, Modi2012, Bera_2018} present in shared multipartite quantum states are shown to be responsible for obtaining advantages in quantum information processing tasks over classical counterparts. 
Among all quantum features, multipartite entanglement emerges as the key ingredient of quantum protocols including quantum networks \cite{Kimble2008} that can transmit both classical and quantum information \cite{bennett1992, brus2004, bruss2006, bennett1993, Murao99, aditi2010}, quantum secret sharing involving a single sender and multiple receivers \cite{Hillery99}, and measurement-based quantum computation \cite{Rausendorf2001, Briegel2001, Briegel2009}. Additionally, it has also been demonstrated that the ground,  thermal or dynamical state of quantum many-body systems \cite{Lewenstein2007, amico2008, DeChiara_2018} possesses  multipartite entanglement which can also be used to identify quantum phenomena \cite{Wei05, Anindya14, stavPRB} 
present in both equilibrium and non equilibrium situations \cite{sachdev_2011, Heyl_2018}.
More importantly, multipartite entangled states can be produced in  laboratories; 
examples include twelve photon genuine multipartite entangled states \cite{Pan18, Pan18b},  trapped ions having twenty qubits \cite{Blatt18},   twenty entangled superconducting qubits  \cite{superconduc20} etc. This opens up the possibility of realizing quantum information protocols involving multipartite entangled states in laboratories. 


\adi{Recently, various methods for generating multipartite entangled states in quantum many-body systems have emerged, notably using  measurements~\cite{zukowski1993, Briegel2001, sen(De)2005, WaltherZeilinger2005, Acin2007, Briegel2009, Cavalcanti2011, Sadhukhan17,
Banerjee2020, Halder2021, Shaji22}
and deterministic quantum gates~\cite{KrausCirac2001, zang2015, sharma2020, Severin2021, Halder2022}, both capable of producing highly genuine multipartite entanglement.
In gate-based schemes within quantum circuits designed for specific algorithms \cite{nielsen_chuang_2010, Barenco1995, Salomaa04}, it is important to characterize unitary operators by their capability to generate multipartite entangled states from product states. The average entangling power of a two-qubit unitary is defined as the mean linear entropy generated when acting on Haar-uniformly distributed product states \cite{Makhlin02, Zanardi00, Zanardi01, wang02, bala09, Bala10, Sudbery05}.}
By exploiting the Cartan decomposition from the theory of Lie groups and algebras, it is possible to obtain an elementary decomposition of an arbitrary unitary operator acting on an arbitrary number of qubits   \cite{KHANEJA200111, Farrokh2004, Farrokh04}. This connection to Cartan decomposition also sheds light on  the geometric structure of the nonlocal unitary operators, which are the only operators responsible to create any entanglement. In particular, Cartan decomposition applied to the Lie algebra $\mathfrak{su}(4)$ gives that the geometric structure of nonlocal gates is a $3$-torus \cite{Whaley03}.  Moreover, estimates of the number of two-qubit gates in near-optimal circuits can be determined by applying Cartan decomposition recursively \cite{Mansky22}. 
However, in contrast to the average entangling power, the maximum entangling power of a fixed unitary operator can also reveal the optimal strength to create an entangled state from the initial product state \cite{cirac01} or a set of specific states having certain amounts of entanglement \cite{leifer01}. Both directions lead to the quantification of a resource theory for operations \cite{Nielsen03}.

The strength of two-qubit quantum gates \cite{Zanardi00, Zanardi01, Galve13} has been evaluated using quantum correlation measures such as entanglement \cite{Horodecki2009} and quantum discord \cite{Bera_2018}. Beyond pure product states and unitaries, their action on mixed states and the entangling power of non-unitary operators have also been explored in bipartite systems \cite{Nonunitary15, Guo14}. Another approach compares unitary operators based on Hamiltonian simulation times \cite{Dur01, Vidal02, Hammerer02}. For symmetric two-qubit unitaries acting on symmetric separable states, explicit expressions for the average entangling power are obtained using Cartan decomposition of semi-simple Lie algebras \cite{symmetric22}. Extending beyond bipartite settings, the entangling capacity of multipartite gates has been investigated via linear entropy when acting on fully separable states \cite{Scott04, Fenner, Linowski_2020}.

\adi{In this work, we depart from the traditional analysis of entangling power in two ways -- first, by altering the set of input product states for optimization, and second, by considering the multipartite entanglement measure for assessing the entanglement content of the output states, obtained after the action of unitary operators under consideration. Specifically, we consider different classes of 
``\(k\)-separable pure states" \cite{blasone2008} for different values of \(k\) as the input states -- for three parties, 
\(3\)-separable (fully separable) and  \(2\)-separable (biseparable) pure states -- ensuring biseparable states exclude  fully separable ones. Precisely, in this multipartite setup, the entangling power of a fixed unitary operator is defined as the maximum genuine multipartite entanglement (GME), measured via the generalized geometric measure (GGM) \cite{wei2003, aditi2010}, generated through a unitary operator from the chosen \(k\)-separable pure inputs, termed as the operator’s {\it genuine multipartite entangling power} (see Fig.~\ref{fig:schematic} for schematic diagram).}

\adi{ We identify classes of unitary operators and input \(N\)-party pure states with vanishing GGM that are capable of producing the maximum possible amount of genuine multipartite entanglement when the unitary operator acts on inputs.} Specifically, we prove that there are unitary operators, namely diagonal and permutation unitaries, in which parameters can be tuned in such a way that the states having maximum GGM   can be produced from some particular kind of \(k\)-separable \adi{pure} states (\(2\leq k \leq N\)).
We also classify the unitary operators based on their capability to generate GME from certain classes of input separable states. In this respect, one might expect that a global unitary operator with \(k\)-separable states as inputs generates more genuine multipartite entanglement as opposed to \(k+1\)-separable states. However, we demonstrate that such an intuition is too simplistic! 
Specifically, there exist unitary operators that can not take advantage of the initial entanglement present in the input states (in certain bipartitions) and produce higher entanglement. Having said that, we determine unitary operators such as diagonal,  permutation, and Haar uniformly generated unitary operators, which indeed provide the benefit of initial entanglement present in bipartition. \adi{We establish these results by considering different classes of eight- and higher-dimensional unitary operators which act on three- and higher-qubit pure input states, respectively, and the optimization is performed over the sets of corresponding separable states. Moreover, we observe that when the inputs are chosen from the set of fully separable states,  the average multipartite entangling power of Haar-uniformly generated unitary operators increases with the increase in the number of parties.  Additionally, our numerical simulations reveal that restricting the optimization to the real space significantly reduces the entanglement produced by random unitary operators compared to the case where input states have complex coefficients, highlighting the substantial role of complex amplitudes in achieving higher multipartite entanglement.}

The paper is organized as follows. In Sec. \ref{sec:definition}, we introduce the notion of genuine multipartite entangling capability of unitary operators based on different classes of separable multipartite inputs and their properties. We identify classes of input states and the corresponding classes of unitary operators which can create maximally genuine multipartite entangled states in Sec. \ref{sec:maximalGGMset}.  Sec. \ref{sec:3qubitsresults} computes entangling power of 
 randomly generated diagonal operators, Haar uniformly produced random unitary operators, and special classes of operators \cite{Farrokh2004} for three-, four- and five-qubit input states. 
The concluding remarks are included in Sec. \ref{sec:conclu}. 

\section{Hierarchies in the entangling power of unitary operators}
\label{sec:definition}

\begin{figure}
\includegraphics[width=1.05\linewidth, height=0.23\textheight]{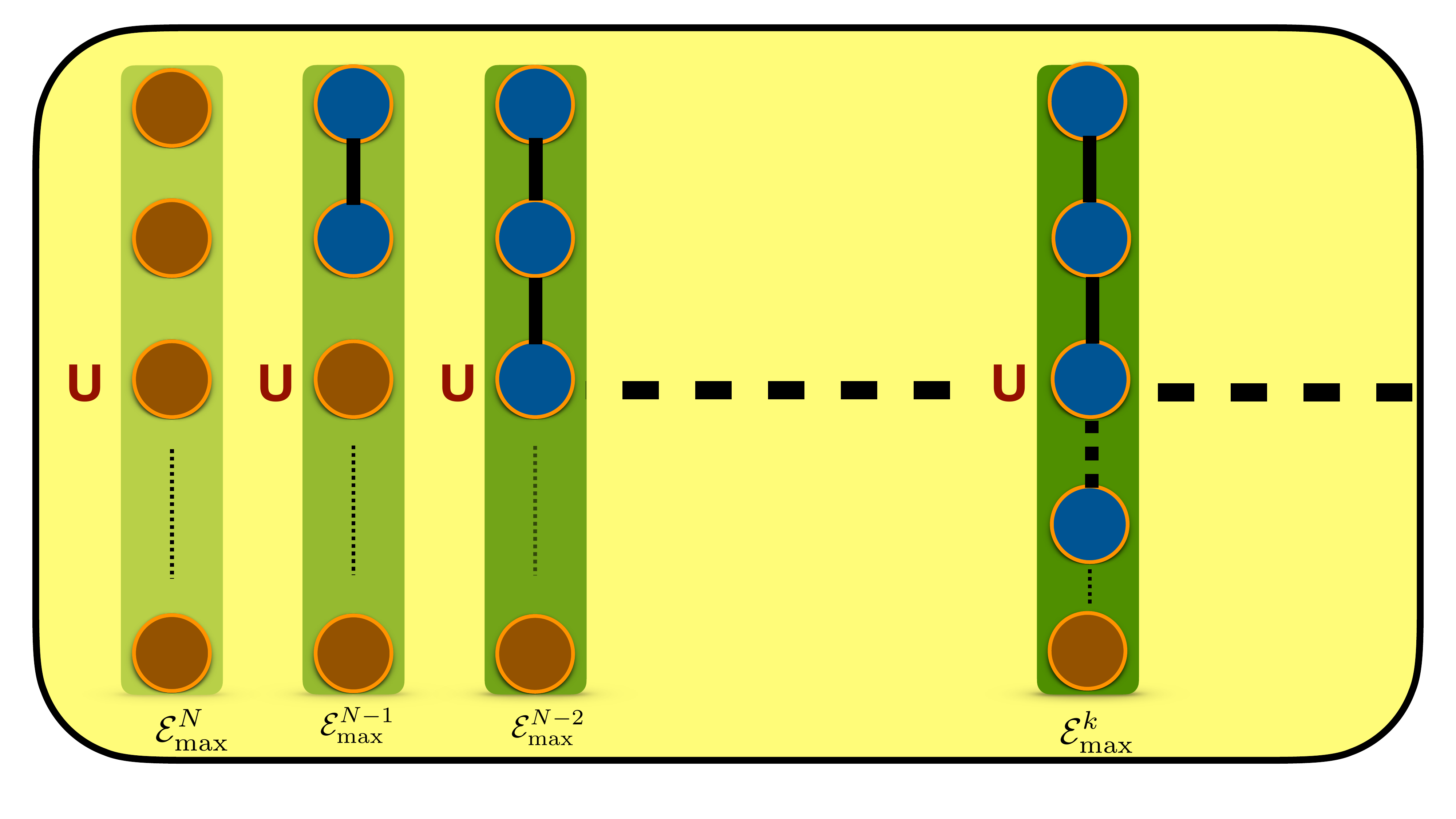}
\caption{(Color online.) A schematic diagram of the notion of hierarchies among entangling powers of a fixed unitary operator, \(U\), based on the inputs chosen.  Each column represents a different set of separable states. From left to right, set of inputs are chosen to be \(N\)-separable (i.e., fully separable) to \(k\)-separable. The parties entangled are marked with blue circles while the brown circles are for the rest. After the action of \(U\) on \(k\)-separable inputs, the maximum genuine multipartite entanglement of the resulting states, \(\mathcal{E}^k_{\max}\) in Eq. (\ref{eq:maindef}),  is computed, which is represented by the vertical green color.  Here, the superscript represents the class of separable states considered for maximization.  Deeper green indicates more expected GME created in the outputs. }
\label{fig:schematic} 
\end{figure}

An \(N\)-party pure state, \(|\psi\rangle\), is said to be \(k\)-separable (\(k=2, \ldots N\)) when it can be written as a product in the \(k\) number of partitions \cite{blasone2008}\footnote{\textcolor{black}{The definition of \(k\)-separability for pure states can be extended to mixed states by using convex roof extension. In particular, a mixed \(N\)-party state $\rho^{k\text{-sep}}$ is said to be \(k\)-separable if it can be written as a convex combination of \(k\)-separable pure states, i.e., 
\(\rho^{k\text{-sep}} = \sum_i p_i |\psi^{k\text{-sep}}_i \rangle \langle \psi^{k\text{-sep}}_i| \). Note that, in this decomposition,  the states, \(|\psi^{k\text{-sep}}_i \rangle \) can be \(k\)-separable in different partitions. In this work, we will mainly concentrate on the set of \(k\text{-sep}\) pure states.}}  Mathematically, 
\begin{eqnarray} 
|\psi^{\text{k-sep}}\rangle_{N} =  \otimes_{m=1}^{k} |\psi_m\rangle. 
\end{eqnarray}
When \(k=N\), the state is called fully separable, while a pure state is called genuinely multipartite entangled when it cannot be written as a product across any bipartition.  Importantly, 
the set of fully separable states, (\(N\)-separable) is a subset of the set of \((N-1)\)-separable states, which, in turn, is a subset of the \((N-2)\)-separable states, and so on, down to the 2-separable (biseparable) states, represented as
\begin{equation}
\mathcal{S}^{N} \subseteq \mathcal{S}^{N-1} \subseteq \cdots \subseteq \mathcal{S}^{k} \subseteq \mathcal{S}^{k-1} \subseteq \cdots \subseteq \mathcal{S}^{2},
\end{equation}
where \(\mathcal{S}^k\) denotes the set of all \(k\)-separable pure states. Note further that with an increasing number of parties, there can be different kinds of entangled states among the \(k\)-separable states themselves. \adi{For example, when \(N=6\), a variety of entanglement structures can arise  within the class of \(3\)-separable states -- (a) each  pair can form a bipartite entangled state, i.e.,  \(|\psi^{3\text{-sep}}\rangle_6 = |\psi_{ij}\rangle \otimes |\psi_{kl} \rangle \otimes |\psi_{mn}\rangle\); (b) three partition contains a genuinely tripartite entangled state, a bipartite entangled state, and a single-qubit state as \(|\psi^{3\text{-sep}}\rangle_6 = |\psi_{ijk} \rangle\otimes |\psi_{lm}\rangle \otimes |\psi_{n}\rangle\); and  (c) it consists of a genuinely quadripartite entangled state accompanied by two single-party  states, i.e., \(|\psi^{3\text{-sep}}\rangle_6 = |\psi_{ijkl}\rangle \otimes |\psi_{m}\rangle \otimes |\psi_{n}\rangle\) with \(\{i,j,k,l,m,n\}\) being any non-identical values between \(1\) to \(6\).}

 {\it Definition of multipartite entangling power.}  A given unitary operator, \(U\), acting on any possible partition of a \(k\)-separable pure state,  can typically produce a genuine multipartite entangled state. When the genuine multipartite entanglement content of the resulting state is maximized over this partition of  \(k\)-separable \adi{pure} states (with a fixed value of \(k>1\)), we call it  \(k\)-maximal entangling power of \(U\) in that partition\footnote{\adi{Instead of pure states, this definition can also be extended to the set of mixed states when both input and output states can be mixed.}}. 
\adi{Specifically, for a given genuine multipartite entanglement measure, \(\mathcal{E}\),  a given unitary operator, \(U\), and a fixed partition \(\alpha\), corresponding to a fixed value of \(k\),  it can be mathematically represented as
\begin{eqnarray}
\mathcal{E}_{\max}^{k} (U) = \max_{\mathbb{S}^{k}} \mathcal{E} (U |\psi^{k\text{-sep},\alpha} \rangle_N), 
\label{eq:maindef}
\end{eqnarray}
where the maximization is carried out over the set \(\mathbb{S}^{k}\) of \(k\)-separable pure states. In this context, it is important to note that in the definition of entangling power of a unitary operator\footnote{Replacing maximum by average in Eq. (\ref{eq:maindef}) {\color{black} where the averaging is performed over the set of \(k\)-separable pure states, we can obtain average entangling power of a unitary operator, \(U\), as 
\begin{eqnarray}
    \overline{\mathcal{E}^k (U)} = \int_{\mathbb{S}^k} \mathcal{E} (U |\psi^{k\text{-sep}} \rangle_N ) d \Omega_N, 
\end{eqnarray}
where \(d\Omega_N\) represents the Haar measure over $\mathbb{S}^k$.}} (see Fig.~\ref{fig:schematic}), no GME measure has been used so far in the literature \cite{Zanardi00, Scott04, Linowski_2020}.  }

{\color{black} {\it Sets for optimization.} To carry out our investigation in a more structured manner, 
we refine the definition of \(k\)-separable states by excluding any states that also belong to higher separability classes, \(\mathcal{S}^{k+i}\) for all \(i \geq 1\). In other words, we restrict our attention to those states that are genuinely \(k\)-separable and not more separable than that. We denote this refined set of \(k\)-separable states by \(\mathbb{S}^k\), defined as 
\[
\mathbb{S}^k := \mathcal{S}^k \setminus \mathcal{S}^{k+1}.
\]
This construction ensures that the sets \(\mathbb{S}^k\) are mutually disjoint, providing a clear and non-overlapping classification of quantum states based on their degree of separability.

To ensure the pure states to be \emph{strictly \(k\)-separable}, we use the following conditions:
\begin{eqnarray}
\nonumber S(\rho_i) = 0 \quad \text{for } 1 \leq i \leq k, \quad \\
\text{and} \quad S(\rho_j) \neq 0 \quad \text{for } k+1 \leq j \leq N,
\label{eq:k_sep}
\end{eqnarray}
where \(S(\rho) = \text{tr}(\rho \log_2 \rho)\) denotes the von Neumann entropy of \(\rho\) and \(\rho_{i(j)}\) represent the reduced density matrices of  \(|\psi^{k\text{-sep}}\rangle_N\). This condition ensures that the state can be decomposed into exactly \(k\) unentangled subsystems, while any further subdivision would necessarily involve entanglement. Consequently, each state can be uniquely identified with one and only one separability class \(\mathbb{S}^k\), eliminating ambiguities arising from overlap between classes of different \(k\).
}
Now, since the number of parties which are entangled in the states of \(\mathbb{S}^k\) is higher than the number of those in \(\mathbb{S}^{k+1}\),  one may expect  
higher production of entanglement from
\(\mathbb{S}^k\), i.e., \(\mathcal{E}_{\max}^{k+1}(U) \leq \mathcal{E}_{\max}^{k}(U)\). We will illustrate that this is not always the case. 
It is important to stress here that all the previous studies considered the entangling power of unitary operators with their action only on the set of fully separable states \cite{Scott04, Linowski_2020}, which obviously does not allow any comparison of our kind. Before studying in detail the entangling power of several classes of unitary operators, let us discuss its properties, in general.

{\it Properties of multipartite entangling power.}  \(\mathcal{E}^k_{\max}(U)\) has the following features. 

(1) It is nonvanishing for all \(U\)s which create GME states as outputs. In other words,   \(\mathcal{E}^k_{\max}(U) =0\) if and only if all the output states have vanishing genuine multipartite entanglement. The upper bound of \(\mathcal{E}^k_{\max} (U)\) is fixed by the choice of the entanglement measure, \(\mathcal{E}\). 

(2) \adi{\(\mathcal{E}^k_{\max}(U)\) remains invariant under local unitary operations, i.e., \(\mathcal{E}^k_{\max} (\otimes_{i=1}^N U_i U) = \mathcal{E}^k_{\max}(U)\). If the entanglement measure chosen is local unitarily equivalent, it can be shown that the multipartite entangling power also remains so.  Suppose before local unitary operation, the optimal state that maximizes entangling power is \(|\psi^{opt}\rangle\), i.e., \(\mathcal{E}^k_{\max} (U) = \mathcal{E} (U |\psi^{opt}\rangle)\) while after local unitary operations, it can go to the other state although entanglement remains intact due to local unitary equivalence of the entanglement measure. Hence the proof. }

In this context also, for a fixed unitary operator \(U\), a similar comparison between {\color{black} \(\overline{\mathcal{E}^k (U)} \) and \(\overline{\mathcal{E}^{k+1} (U)} \) can also be investigated. In this work, we restrict ourselves to \(\mathcal{E}^{k}_{\max} (U)\).  }


It is important to stress here that in order to build the quantum circuits composed of higher-dimensional unitary operators considered in this work, one can decompose them in terms of single- and two-qubit operations  \cite{Barenco1995}. In particular, the operators we consider here are all in $SU(2^N)$, which is a semi-simple Lie group. Any such operator can be decomposed into a product of elements of $SU(2)$ and $SU(4)$ following a method described in Ref. \cite{KHANEJA200111}. This method is based on successive application of the Cartan decomposition of the Lie group $SU(2^N)$, which also uses geometric facts about $\frac{SU(2^N)}{SU(2^{N-1})\otimes SU(2^{N-1})\otimes U(1)}$. The successive way of writing an $N$-dimensional operator in low dimensional operations also provides an algorithmic approach to the problem.

{\it Quantification of GME.}  We use generalized geometric measure to compute genuine multipartite entanglement content of the output states \cite{aditi2010}. For an \(N\)-party state, \(|\chi\rangle\),  it is defined as
\(G (|\chi\rangle) = 1 - \max\limits_{|\phi\rangle \in \mathcal{D}} |\langle \phi | \chi \rangle|^2 \), where \(\mathcal{D}\) denotes the set of all non-genuinely multipartite entangled states. It has been proven that the maximization involved in GGM can be simplified by using Schmidt coefficients of \(|\psi \rangle\) in different bipartitions \cite{aditi2010}. Note that \(G(|\chi \rangle) = 0\) for any  \(k\)-separable state \(|\chi\rangle\), in particular for any state in \(\mathbb{S}^k\) (\(k>1\)). In this work, \adi{we will compute the maximal GGM as \(\mathcal{E}\) in Eq. (\ref{eq:maindef})} produced by the action of \(U\) on the input states from the set \(\mathbb{S}^k\).  \adi{Instead of \(\mathcal{E}^k_{\max}\), we henceforth denote the {\it genuine multipartite entangling power} as \(G^k_{\max}\) for different \(k\) values of initial separable states or alternatively \(G_{\max}\).} 


\section{Unitary operators with Maximal multipartite entangling power and their Inputs}
\label{sec:maximalGGMset}

Towards characterizing unitary operators (quantum gates) in terms of their entangling power,  we identify the operators that create multipartite resulting states having {\it maximum} GGM. In the process, we also determine the optimal inputs which give us such outputs. Further, these investigations provide a deterministic method to produce states having maximum GGM useful in quantum information processing tasks.  



\subsubsection*{Setting up the stage for three-qubits}

Let us first describe how one can determine both optimal inputs and the corresponding outputs for a fixed unitary operator which is capable of producing maximum GGM for three-qubits. 

A pure three-qubit separable state  can be either fully separable, 
\[|\psi^{3\text{-sep}} \rangle_3 = |\psi_1\rangle \otimes |\psi_2\rangle \otimes |\psi_3\rangle = \otimes_{i=1}^3 (\cos \theta_i |0\rangle + \sin \theta_i e^{i \xi_i} |1\rangle), \] 
with \(0\leq \theta_i \leq \pi/2\) and  \(0\leq \xi_i \leq 2\pi\)
or a biseparable state  \cite{Durvidal2002}   of the form 
\begin{eqnarray}
|\psi^{2\text{-sep}} \rangle_3 &= &|\psi_i\rangle \otimes |\psi_{jk}\rangle, \, i \neq j \neq k,
\end{eqnarray} 
where \(|\psi_{jk}\rangle\) is a two-qubit entangled state. For example, considering \(i=1\),  \(|\psi^{2\text{-sep}} \rangle_3\)  can be parametrized as  \((\cos \theta'_1 |0\rangle + \sin \theta'_1 e^{i \xi'_1} |1\rangle) \otimes \sum_{j,k=0}^1 a_{jk} |jk \rangle\) with
  \(a_{00} = \cos \theta'_2, a_{01} = e^{i\xi'_1}\sin \theta'_2 \cos \theta'_3, a_{10} = e^{i\xi'_2} \sin \theta'_2 \sin \theta'_3 \cos \theta'_4, a_{11} = e^{i\xi'_3} \sin \theta'_2 \sin \theta'_3 \sin \theta'_4\)  in terms of higher dimensional spherical polar coordinates and phases. Due to the constraint that sets of biseparable (\(\mathbb{S}^2\)) and fully separable states  (\(\mathbb{S}^3\)) are disjoint,  \(\theta'_i\)s have to be chosen in such a way that the bipartite states in \(\mathbb{S}^2\)  \adi{always} have nonvanishing entanglement \cite{Horodecki2009}.

Let us identify the unitary operators on eight-dimensional complex Hilbert space, \(\mathbb{C}^8\), denoted by \(U(8)\),  which act on the  states chosen from \(\mathbb{S}^2\) or \(\mathbb{S}^3\) such that 
the resulting states after maximization over the inputs reach the maximal value of GGM, i.e., \(0.5\)\footnote{\adi{The maximum GGM for \(N\)-qubit state is \(0.5\) since the squared Schmidt coefficient in the definition of GGM is in \([0.5, 1]\).}}. In other words, our aim is to find out whether there exists any  biseparable (optimal) input, $|\psi^{2\text{-sep}}_{opt}\rangle_3$,  for a fixed element of $U(8)$, again denoted by $U(8)$, so that
\begin{eqnarray}
G_{\max}^2(U(8)) \equiv G(U(8)|\psi^{2\text{-sep}}_{opt} \rangle_3) =0.5, 
\end{eqnarray}
  and in the case of fully separable ones, 
 \begin{eqnarray}
G_{\max}^3(U(8)) \equiv G(U(8))|\psi^{3\text{-sep}}_{opt} \rangle_3) =0.5.
 \end{eqnarray}
We will demonstrate that, indeed,  there are unitary operators and corresponding inputs for which both the above equations hold true. However, there exist eight-dimensional unitary operators for which only one of the above equations can be satisfied. In what follows, we will encounter such unitary operators. 
 
 

 \subsection{Multipartite entangling power of diagonal unitary operators} 
 \label{subsec:entpowerdiag}

Let us illustrate the significance of inputs in the context of deciding the multipartite entangling power of diagonal unitary operators. 
A diagonal unitary operator on an eight-dimensional space can be written as 
\begin{eqnarray}
    U_D^{gen} = \mbox{diag}(e^{i\phi_1}, e^{i\phi_2}, e^{i\phi_3}, e^{i\phi_4}, e^{i\phi_5}, e^{i\phi_6},  e^{i\phi_7}, e^{i\phi_8}), 
\end{eqnarray}
where \(\phi_j \in [0, 2 \pi)\) \cite{Fenner} (\(j=1, \ldots, 8\)). 
 
We concentrate here on the diagonal unitary operators with  \(\phi_j =0\) except \(\phi_8\), thereby reducing to the case \(U_D= \mbox{diag}(1,1,1,1,1,1, 1, e^{i\phi})\). The entangling power of general diagonal unitary operators with arbitrary \(\phi_j\)s will be investigated in the next section.

After application of such an operator, $U_D$, on general fully separable or biseparable states, the resulting states  read as \(\sum_{i,j,k=0}^1 a_{ijk} e^{i \phi_{ijk}}|ijk\rangle \), where \(e^{i \phi_{ijk}} =1\) except \(e^{i \phi_{111}} \equiv e^{i \phi}\) and  \(a_{ijk}\)s are functions of  \(\theta_i\)s, \(\theta'_i\)s, \(\xi_i\)s and \(\xi'_i\)s.  To obtain GGM  for this resulting three-qubit state, we need to calculate all the single-qubit reduced density matrices and their eigenvalues. 
We first notice that during optimization, \(\xi_i\)s and \(\xi'_i\)s are not playing any role and therefore, we set them to be zero, i.e., in the optimization process, phases present in \(|\psi^{2\text{-sep}}\rangle_3\) and \(|\psi^{3\text{-sep}}\rangle_3\) are not involved {\color{black}(See Appendix \ref{sec:bi_entagling_power} and Appendix \ref{sec:full_entagling_power} )}. This implies the coefficients $a_{ijk}$ are now real-valued functions.

We first consider applying the operators of elements of \(\mathbb{S}^2\). After application of \(U_D\) on an arbitrary biseparable state, the output takes the form, 
\begin{eqnarray}
   |\psi_{out}\rangle &\equiv& U_D|\psi^{2\text{-sep}}\rangle_3 \nonumber\\ 
   &= & 
   \cos\theta'_1 \cos \theta'_2 |000\rangle +  \cos \theta'_1 \cos \theta'_3  \sin \theta'_2 |001\rangle \nonumber\\
   & + & \ldots  +
e^{i\phi}
   \sin \theta'_1 \sin \theta'_2 \sin \theta'_3 \sin \theta'_4 |111\rangle,
   \label{equ:full_out_diag_one}
\end{eqnarray} 
where only the last term has \(\phi\)-dependence. In order to compute GGM for these states,
the corresponding single-site local density matrices can be obtained, which are functions of the form \(G (\theta'_1,\ldots ,\theta'_4, \phi)\). For a fixed \(\phi\), after optimizing over \(\theta'_i\)s, the  optimal input biseparable state turns out to be \begin{eqnarray}
    |\psi^{2\text{-sep}}_{opt}\rangle_3 = \frac{1}{\sqrt{2}} (|0\rangle + |1\rangle) \frac{1}{\sqrt{2}}(|00\rangle + |11\rangle),
\end{eqnarray}
which leads to the resulting state, 
 \(|\psi_{out}\rangle= \frac{1}{2}(|000\rangle + |011\rangle + |100\rangle + e^{i\phi} |111\rangle)\) \adi{(see Appendix \ref{sec:bi_entagling_power} for details)}. 
 When \(\phi =\pi\),  all single-qubit reduced density matrices  of \(|\psi_{out}\rangle \) are maximally mixed. Therefore,  the GGM of the output state achieves its maximal value, i.e., 
 \begin{eqnarray}
     G_{\max}^2 (U_D(\phi = \pi)) \equiv G(U_D(\phi = \pi)|\psi^{2\text{-sep}}_{opt}\rangle_3) = 0.5.
 \end{eqnarray}
 In general, for arbitrary \(\phi\), all the local density matrices of  \(|\psi_{out}\rangle\) are \(\mathbb{I}/2\) except the first one which is given by {\color{black}\(\frac12\big(|0\rangle \langle 0|+e^{-i\frac{\phi}{2}}\cos\frac{\phi}{2} |0\rangle \langle 1| + e^{i\frac{\phi}{2}}\cos\frac{\phi}{2} |1\rangle \langle 0| +|1\rangle \langle 1|\big)\).} 
Therefore, in this case, we obtain
 \begin{eqnarray}
   G_{\max}^2 (U_D) \equiv  G (U_D|\psi^{2\text{-sep}}_{opt}\rangle_3) = \sin^2 \frac{\phi}{4}, 
 \end{eqnarray}
 when \(0\leq \phi \leq \pi\).
 From the above expression, it is clear that $G_{\max}^2$ reaches its maximum value only for \(\phi =\pi\) and trivially vanishes at \(\phi =0\).

\begin{figure}
\includegraphics[width=1.0\linewidth]{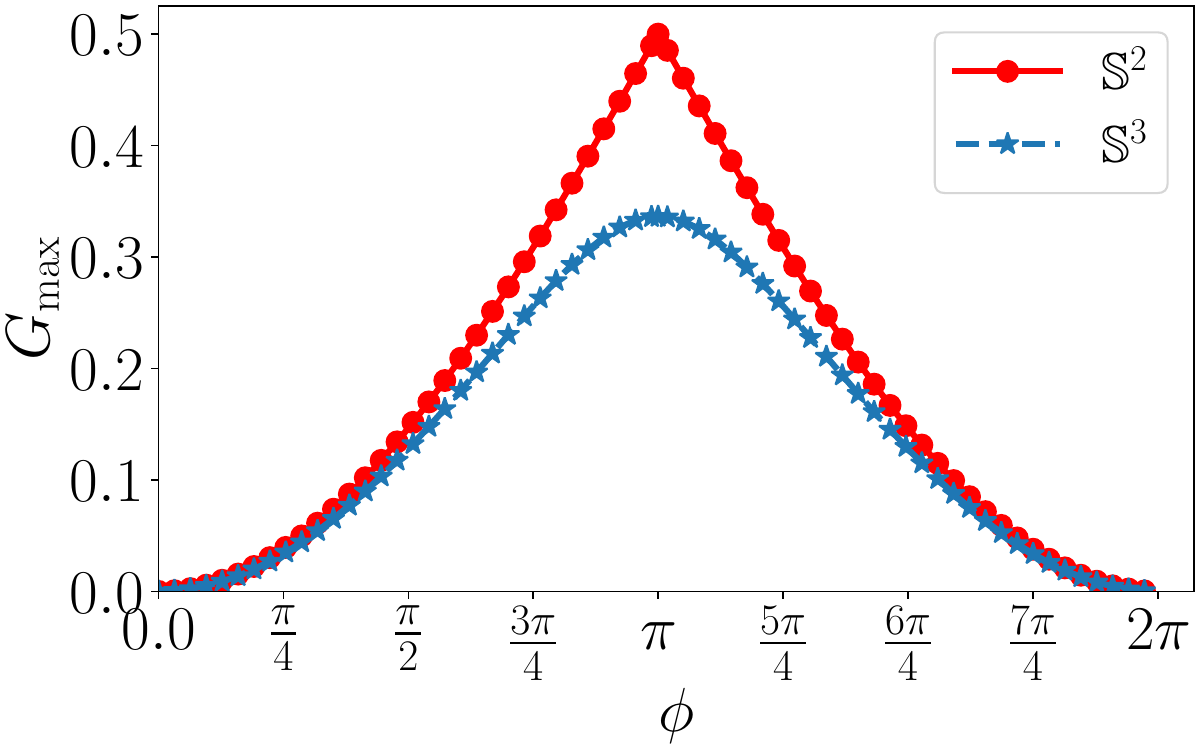}
\caption{(Color online.) {\bf Enhancement in entanglement generation with initial entanglement.}  Biseparable inputs are better than the fully separable ones when a special diagonal unitary matrix, \(U_D= \mbox{diag}(1,1,1,1,1,1, 1, e^{i\phi})\) with \(\phi \in (0, 2 \pi)\) \cite{Fenner} acts on three-party input states. The maximum GGM produced (ordinate) against \(\phi\) (abscissa). The optimization is performed over the set of biseparable (circles) and fully separable (stars) states, denoted as \(\mathbb{S}^3\) and \(\mathbb{S}^2\) respectively, to obtain \(G_{\max}\). 
Both the axes are dimensionless.     }
\label{fig:diagonalsp} 
\end{figure}



 On the other hand, if one applies \(U_D\) to the set of  fully separable states, \(\otimes_{i=1}^3(\cos \theta_i |0\rangle + \sin \theta_i |1\rangle)\), the output states take the form,  
 \begin{eqnarray}
    U_D|\psi^{3\text{-sep}} \rangle_3 &= & \Pi_{i=1}^3\cos \theta_i |000\rangle + \cos \theta_1 \cos \theta_2 \sin \theta_3 |001\rangle \nonumber \\
    &+& \ldots + e^{i \phi} \sin \theta_1 \sin \theta_2 \sin\theta_3
      |111\rangle,
 \end{eqnarray}
 where only the coefficient of \(|111\rangle\) depends on \(\phi\).  Again finding all the reduced density matrices and performing maximization over \(\theta_i\)s,  we find that maximum eigenvalues of all the single-party reduced density matrices exceed \(1/2\), i.e., \(G_{\max}^3(U_D)\) never reaches the maximum value of \(0.5\) for any values of \(\phi\).  For example, when \(\phi=\pi\), \(G_{\max}^3(U_D(\phi=\pi)) =0.34\) \adi{(see Fig. \ref{fig:diagonalsp} and Appendix \ref{sec:full_entagling_power})}.  This diagonal unitary operator clearly demonstrates how initial entanglement may assist in the generation of high GME states, although it is not true in general for an arbitrary unitary operator of \(U(8)\), as we will show below. 
 

 { \textbf{Remark}.} {\color{black} It is known that the application of a four-dimensional diagonal unitary, \(U^4_D(\phi=\pi)\), on nearest-neighbor sites of a suitably chosen fully separable state can generate a three-qubit cluster state, which possesses maximal genuine multipartite entanglement and is useful for one-way quantum computation~\cite{Rausendorf2001}. In particular,  considering the operator \(U(8) = \big(U^4_D(\phi=\pi) \otimes I_2 \big)\big(I_2 \otimes U^4_D(\phi=\pi)\big)\), the maximal GGM value of \(0.5\) can  be achieved from a fully separable input state, \(\frac{1}{\sqrt{8}}(\ket{0}+\ket{1})^{\otimes 3}\).  The unitary, in this case, takes the form as \(\mbox{diag}(1,1,1,-1,1,1,-1,1)\) which is different from the one considered above. Our results show that applying a single eight-dimensional diagonal unitary of the form \(U_D\) on any fully separable three-qubit state can never produce a maximum GGM. This limitation persists even when the dimension is increased beyond eight.} 



\subsubsection*{Action of higher dimensional unitaries on multiqubit inputs}

Let us now examine the action of a class of diagonal unitary operators in the sixteen-dimensional complex Hilbert space, given by \(U_{\phi}^{16}= \mbox{diag}(1,\ldots, e^{i\phi})\). In a four-party system,  \(2\text{-sep}\) states of \(\mathbb{S}^2\) can be written as
\begin{eqnarray}
   |\psi^{2\text{-sep}} \rangle_4 =&&|\psi_i\rangle \otimes |\psi_{jkl}^{gen}\rangle,  \text{or}\nonumber\\
 |\psi^{2\text{-sep}} \rangle_4 =&&|\psi_{ij}\rangle \otimes |\psi_{kl}\rangle; (i, j, k, l \text{ are different}),
\label{eq:4qbisep}
\end{eqnarray}
 where \(|\psi_{jkl}^{gen}\rangle\) represents  genuine   tripartite entangled states, while \(|\psi_{\ast\ast}\rangle\)s are the bipartite entangled states. When  \(U_{\phi}^{16}\) acts on the initial four-party states, the resulting states take the form, \(|\psi^{out}\rangle = \sum_{i,j,k,l =0}^1 a_{ijkl} e^{i \phi_{ijkl}} | ijkl\rangle\) with \( e^{i \phi_{ijkl}} =1\) except \(e^{i\phi_{1111}} =e^{i \phi}\), i.e., after its action on the states, the coefficient of \(|1111\rangle\) only depends on  \(\phi\) present in \(U_{\phi}^{16}\) and  \(a_{ijkl}\)s, which depend on the input states, are typically products of cosine and sine functions involved e.g., in \(|\psi^{2\text{-sep}}\rangle_4\).  As in the case of three-qubits, the maximization process reveals that the optimal inputs are either 
 \( |\psi_{opt}^{2\text{-sep}}\rangle_4 = \frac{1}{\sqrt{2}} (|0\rangle + |1\rangle) \frac{1}{\sqrt{2}}(|000\rangle + |111\rangle\)    or \(|\psi^{2\text{-sep}}_{opt}\rangle_4 = \frac{1}{\sqrt{2}} (|00\rangle + |11\rangle) \frac{1}{\sqrt{2}}(|00\rangle + |11\rangle\). Therefore, the multipartite entangling power of \(U_{\phi}^{16}\) turns out to be
 \begin{eqnarray}
      G_{\max}^2(U_{\phi}^{16}) \equiv  G (U_{\phi}^{16}|\psi^{2\text{-sep}}_{opt}\rangle_4) = \min [\sin^2 \frac{\phi}{4}, \cos^2 \frac{\phi}{4}]
      \label{eq:entpower4}
 \end{eqnarray}
 since all the reduced density matrices are either \(\frac{\mathbb{I}}{2}\) or their eigenvalues are  \(\frac{1 \pm \cos\frac{\phi}{2}}{2}\). 
It again attains maximum value, \(0.5\) with \(\phi = \pi\). The numerical simulations performed on the set of \(k-\)separable (\(k>2\)) inputs, i.e., over \(\mathbb{S}^3\) and  \(\mathbb{S}^4\), exhibit that  \(U_{\phi}^{16}\) cannot produce output states having maximal GGM from the set of \(3\)- and \(4\)-separable input states. 

 The above analysis clearly gives an indication that multipartite entangling power of the class of diagonal unitary operators in \(\mathbb{C}^{2^N}\) of the form \(U_{\phi}^{2^N} = \text{diag}(1,1,\ldots, e^{i\phi})\)  always achieve maximum value of \(0.5\) with input \(|\psi^{2\text{-sep}}\rangle_N = \frac{1}{\sqrt{2}} (|0\rangle + |1\rangle) \frac{1}{\sqrt{2}}(|0\ldots 0\rangle + |1\ldots 1\rangle\). Moreover, starting with any \(N\)-party input state of the form,  \(|\psi^{2\text{-sep}}\rangle_N = \frac{1}{\sqrt{2}} (|0\rangle^{\otimes k} + |1\rangle^{\otimes k}) \frac{1}{\sqrt{2}}(|0\rangle^{\otimes N-k} + |1\rangle^{\otimes N-k})\) for different \(k\), we obtain \(G_{\max}^2(U_{\phi = \pi}^{2^N}) =0.5\) while for arbitrary \(\phi\), the formula for GGM is as given in Eq. (\ref{eq:entpower4}).  
 In this manner, we identify both, a class of diagonal unitary operators and their optimal inputs, which are capable of producing maximum GGM in a given dimension. 
 

 \subsection{Entangling power for a set of permutation unitary operators }
\label{subsec:entpermut}


 Let us concentrate on the set of permutation unitary operators in the eight-dimensional space.  In this case, unlike diagonal unitaries, we will illustrate that both fully separable and biseparable states can create outputs having maximum GGM.  Let us first fix a product basis in \(\mathbb{C}^8\), \(\mathcal{E}= \{\mathcal{E}_{0} \equiv |\alpha_1 \alpha_2 \alpha_3 \rangle, \mathcal{E}_{1} \equiv |\alpha_1 \alpha_2 \alpha_3^{\perp} \rangle, \ldots, \mathcal{E}_{8} \equiv |\alpha_1^{\perp} \alpha_2^{\perp} \alpha_3^{\perp} \rangle\} \) {\color{black}with \(\{\ket{\alpha_i},\ket{\alpha_i^\perp}\}\)  (\(i=1,2,3\)) being the basis of the respective Hilbert spaces}. {\color{black}We will consider the permutation operators of the form \(\Pi_{(i,j)}\) where \(i,j (= 1, \ldots, 8)\), called transpositions.  For example, it can be taken to be a computational basis, \(\{\ket{0},\ket{1}\}\).} They act by permuting the basis elements  \(\mathcal{E}_i\) and  \(\mathcal{E}_j\) while keeping other basis elements unchanged. In eight-dimensional space, \(28\) such permutation operators exist for a given basis.  Our aim is to identify all such permutation operators that produce maximally entangled state {\color{black}in terms of GGM}, i.e.,  \(G(*) = 0.5\). {\color{black}
To obtain maximum GGM, we first notice that the tripartite output state should be symmetric under permutation of parties and all  its reduced density matrices should be maximally mixed as a absolutely maximally entangled state has maximal entropy in each of its subsystems \cite{Brown2005,Borras2007,Facchi2008}.} {\color{black} E.g., the three-qubit state of the form \(|\psi\rangle = \frac{1}{\sqrt{2}}(|\alpha_1 \alpha_2 \alpha_3 \rangle + |\alpha_1^{\perp} \alpha_2^{\perp} \alpha_3^{\perp} \rangle) \)  achieves maximum GGM which is known as  Greenberger-Horne-Zeilinger(GHZ)  state} \cite{greenberger1989}. 

 {\bf Fully separable inputs:} Among twenty-eight transpositions mentioned above, we identify that twelve of them can produce maximum GGM after acting on fully separable states. 
 Specifically,  we obtain
 \begin{eqnarray}
 G^3_{\max}((i,j)) \equiv G ((i,j) |\psi_{opt}^{3\text{-sep}}\rangle) =0.5. 
 \end{eqnarray}  
 The list of transposition unitary operators and their respective optimal fully separable inputs are given in Table \ref{tab:permutfullysep}. 
 Without loss of generality, we denote {\color{black}\(|\alpha_i\rangle\)
and \(|\alpha_i^{\perp}\rangle\) as \(|0\rangle\) and \(|1\rangle\)} respectively. The above table identifies the transposition operators which have maximum multipartite entangling power if the input set of optimization is restricted to fully separable states. On the other hand, if the optimization is carried out with the set of biseparable states as inputs,  the maximum entangling power of the above twelve transposition operators cannot be achieved.

  

  \begin{table}[]
      \centering
      \begin{tabular}{|c|c|c|}
      \hline
      No. & Unitary & Fully Separable Inputs \\ 
          \hline 
        $1 $ & $\Pi_{(1,4)}$  & $\frac{1}{\sqrt{2}}(|011\rangle+|111\rangle)$   \\
         $2 $ & $\Pi_{(1,6)}$  & $\frac{1}{\sqrt{2}}(|101\rangle+|111\rangle)$ \\
         $3 $ & $\Pi_{(1,7)}$  & $\frac{1}{\sqrt{2}}(|110\rangle+|111\rangle)$\\
        $4 $ & $\Pi_{(2,3)}$  & $\frac{1}{\sqrt{2}}(|001\rangle+|101\rangle)$\\
          $5 $ & $\Pi_{(2,5)}$  & $\frac{1}{\sqrt{2}}(|100\rangle+|110\rangle)$\\
        $6 $ &  $\Pi_{(2,8)}$  & $\frac{1}{\sqrt{2}}(|110\rangle+|111\rangle)$\\
         $7 $ & $\Pi_{(3,5)}$  & $\frac{1}{\sqrt{2}}(|100\rangle+|101\rangle)$   \\
         $8 $ & $\Pi_{(3,8)}$  & $\frac{1}{\sqrt{2}}(|101\rangle+|111\rangle)$ \\
         $9 $ & $\Pi_{(4,6)}$  & $\frac{1}{\sqrt{2}}(|100\rangle+|101\rangle)$\\
         $10 $ & $\Pi_{(4,7)}$  & $\frac{1}{\sqrt{2}}(|100\rangle+|110\rangle)$\\
          $11 $ & $\Pi_{(5,8)}$  & $\frac{1}{\sqrt{2}}(|000\rangle+|100\rangle)$\\
        $12 $ &  $\Pi_{(6,7)}$  & $\frac{1}{\sqrt{2}}(|001\rangle+|101\rangle)$
         \\  
         \hline
      \end{tabular}
      \caption{The permutation operators and their corresponding fully separable inputs which maximize GGM are listed in the second and third columns, respectively. Note that one of the parties in the fully separable inputs is in \(|+\rangle =\frac{1}{\sqrt{2}}(|0\rangle + |1\rangle)\) state which has (nonvanishing) maximum coherence \cite{coherenceRMP}.}
      \label{tab:permutfullysep}
  \end{table}


  {\color{black}{\bf Biseparable inputs. } 
 Let us now change the set of input states to be the set of biseparable states. Again, we determine the twelve transposition operators \(\Pi_{(i,j)}\) which achieve the maximal entangling power, \(0.5\) (see Table \ref{tab:permutbisep}) from biseparable inputs. To generate maximum GGM from the transposition operators, we find that one of the reduced two-party density matrices of the optimal biseparable input state has to be maximally entangled. We also observe that when a transposition \(\Pi_{(i,j)}\) of Table \ref{tab:permutbisep} acts on fully separable inputs,  the GGM of the resulting states are always the same, which turn out to be \(0.33\) and never reach the maximum possible value of $0.5$. In other words,  to have maximum  GGM  upon action of these transposition operators, it requires the input states to have entanglement in some bipartition.    Like the diagonal unitary operators,   we notice that bipartite entanglement can sometimes provide advantages for a set of transposition operators, although it is not true, in general.}

 {\bf Remark 1.} The optimal fully separable input states shown in Table \ref{tab:permutfullysep} are  not {\it unique}. {\color{black} For example, when the permutation unitary \(\Pi_{(1,4)}\) acts on the fully separable states \(\frac{1}{\sqrt{2}}(\ket{011}+\ket{111})\) or \(\frac{1}{\sqrt{2}}(\ket{000}+\ket{100})\), the resulting states also attain the maximum GGM. Similarly, in the case of biseparable states, the input state corresponding to the permutation unitary  \(\Pi_{(1,2)}\), are \(\frac{1}{\sqrt{2}}(\ket{000}+\ket{110})\) or \(\frac{1}{\sqrt{2}}(\ket{001}+\ket{111})\). Moreover, permutation unitary is a restricted class of entangling unitary operator which only transforms a class of fully separable or biseparable input states to maximum GME states.}  
  
 { \color{black}{\bf Remark 2.} To generate maximum GGM, one of the parties in the initial optimal state should have maximum coherence \cite{coherenceRMP}, ensuring superpositions across  computational basis states, which under the action of permutation operators can lead to
maximal interference effects, thus maximizing the GGM.}

 { \color{black}{ \textbf{Remark 3.}}  The unitaries involved exhibit symmetries that effectively change the basis such that the initially separable product state evolves into a maximally entangled state. This transformation relies on the specific structure of the unitaries and their action on the input state, which aligns with the entangling power of the unitary operator. However, in the case of relabeling the three qubits, the results are not invariant under arbitrary permutations. To obtain a maximally entangled state from a fully separable or biseparable initial state, one must apply a suitable permutation (or relabeling) along with an appropriate unitary that preserves the entangling structure. Not all relabeling will maintain this condition; in fact, some permutations may require additional entangling operations to reach the same degree of entanglement.}


 



\begin{table}[]
      \centering
      \begin{tabular}{|c|c|c|}
      \hline
      No. & Unitary & Biseparable Inputs \\ 
          \hline 
        $1 $ & $\Pi_{(1,2)}$  & $\frac{1}{\sqrt{2}}(|000\rangle+|110\rangle)$   \\
         $2 $ & $\Pi_{(1,3)}$  & $\frac{1}{\sqrt{2}}(|000\rangle+|101\rangle)$ \\
         $3 $ & $\Pi_{(1,5)}$  & $\frac{1}{\sqrt{2}}(|000\rangle+|011\rangle)$\\
        $4 $ & $\Pi_{(2,4)}$  & $\frac{1}{\sqrt{2}}(|001\rangle+|100\rangle)$\\
          $5 $ & $\Pi_{(2,6)}$  & $\frac{1}{\sqrt{2}}(|001\rangle+|010\rangle)$\\
        $6 $ &  $\Pi_{(3,4)}$  & $\frac{1}{\sqrt{2}}(|010\rangle+|100\rangle)$\\
         $7 $ & $\Pi_{(3,7)}$  & $\frac{1}{\sqrt{2}}(|001\rangle+|010\rangle)$   \\
         $8 $ & $\Pi_{(4,8)}$  & $\frac{1}{\sqrt{2}}(|000\rangle+|011\rangle)$ \\
         $9 $ & $\Pi_{(5,6)}$  & $\frac{1}{\sqrt{2}}(|010\rangle+|100\rangle)$\\
         $10 $ & $\Pi_{(5,7)}$  & $\frac{1}{\sqrt{2}}(|001\rangle+|100\rangle)$\\
          $11 $ & $\Pi_{(6,8)}$  & $\frac{1}{\sqrt{2}}(|000\rangle+|101\rangle)$\\
        $12 $ &  $\Pi_{(7,8)}$  & $\frac{1}{\sqrt{2}}(|001\rangle+|111\rangle)$
         \\  
         \hline
      \end{tabular}
      \caption{The second and third columns correspond to the transformation operators and their corresponding biseparable inputs, respectively which lead to maximum GGM, \(0.5\). In this case, two-qubit reduced state is always maximally entangled, \(\frac{1}{\sqrt{2}}(|00\rangle + |11\rangle)\). Note that the entangling power of the above unitaries cannot reach maximum when optimizing over the set of fully separable states. }
      \label{tab:permutbisep}
  \end{table}

 \subsection{ Entangling power of special classes of unitary operators} 
 \label{subsec:entspecial}

\adi{Let us consider a specific class of unitary operators in \(\mathbb{C}^8\). It was shown that in the case of three-qubits, any unitary operator can be decomposed into a number of two-qubit and a single qubit,  along with two special kinds of three-qubit unitary operators~\cite{Farrokh04}, } given by
\begin{eqnarray}
    U_{sp}^1 =  \exp [i  (\sum_{i=x,y, z} J_i \sigma_i \otimes \sigma_i \otimes \sigma_z )],
    \label{eq:spUFarook}
\end{eqnarray} 
and 
\begin{eqnarray}
    U_{sp}^2 = \exp [i(\sum_{i=x,y,z} J_i \sigma_i \otimes \sigma_i \otimes \sigma_x 
+J_4 \mathbb{I} \otimes \mathbb{I} \otimes \sigma_x)].
    \label{eq:spUFa}
\end{eqnarray} 
Here \(\sigma_i\)s (\(i=x, y, z\)) are the Pauli matrices,  and \(J_i\)s are constants. {\color{black} This structure mirrors the interaction Hamiltonian found in systems with collective couplings, such as dipolar or Ising-like models \cite{PhysRevA.62.022311,PhysRevLett.81.3108,Islam2013,Zeiher2016}, and is capable of generating strong tripartite entanglement from initially separable states. In the case of \(U^1_{\mathrm{sp}}\), the presence of \(\sigma_z\) on the third qubit acts as a control that conditions the collective interaction of the first two qubits, thereby enabling the generation of genuine tripartite entanglement through phase correlations.

On the other hand, the 
\(
U^2_{\mathrm{sp}}\)
introduces an additional interaction term of the form \(\mathbb{I} \otimes \mathbb{I} \otimes \sigma_x\), along with a fixed Pauli-\(\sigma_x\) on the third qubit, which adds asymmetry to the coupling and mimics control-like operations or driven evolution scenarios. The term \(\sum_i J_i\, \sigma_i \otimes \sigma_i \otimes \sigma_x\) still enforces pairwise alignment across qubits, while the additional single-qubit \(\sigma_x\) rotation on the third qubit introduces further tunability. Altogether, these unitaries serve as minimal but tunable blocks capable of entangling three-qubit systems in nontrivial ways. Their parametric simplicity makes them both analytically tractable and experimentally realizable in systems such as superconducting qubits \cite{ Shainline2017, PhysRevApplied.7.054025, melanson2019}, trapped ions \cite{PhysRevLett.129.063603}, or cold atoms \cite{PhysRevA.89.053619}, where controlled multiqubit interactions are increasingly accessible. Thus, their study is well-motivated beyond the mathematical decomposition ~\cite{Vatan2004}.}

By examining optimizations over the set of fully separable states,  we find that  \(|+\rangle^{\otimes 3}\) is an optimal input for the operators \(U^1_{sp}\) with \((J_y -J_z) = n\pi/4, \,  (n=\pm 1, \pm 3, \pm 5, \ldots)\) which leads to maximum GGM. 
In particular, we obtain
\begin{eqnarray}
&& |\psi_{out} \rangle \equiv  U^1_{sp}|+\rangle^{\otimes 3} = \frac{1}{2 \sqrt{2}}[a_1(|000\rangle + |110\rangle) \nonumber\\
 &&+ \overline{a_1} (|001\rangle + |111\rangle) 
    + a_3(|010\rangle + |100\rangle)  \nonumber\\
    &&+ \overline{a_3} (|011\rangle + |101\rangle)],\nonumber \\
\end{eqnarray}
where \(a_1 = \exp[i (J_x-J_y + J_z)]\) and \(a_3 = \exp[i (J_x + J_y - J_z)]\). Clearly, the GGM for this state turns out to be
\begin{eqnarray}
    G(|\psi_{out}\rangle) = 1 -\max[\frac{1}{2} (1 \pm \cos 2(J_y-J_z))],
\end{eqnarray}
which imposes the above mentioned restriction. Thus, we have identified a special subset \(\tilde{U}^1_{sp} \subset {U}^1_{sp} \) given by $\tilde{U}^1_{sp}:=\{U \in {U}^1_{sp} \,|\, J_y -J_z = n\pi/4, \,  n=\pm 1, \pm 3, \pm 5, \ldots \}$.

In contrast, by numerical simulation, we find that the elements of $\tilde{U}^1_{sp}$ do not always produce maximal GGM from the biseparbale states. 
For example, when the operator in \(\tilde{U}_{sp}^1\) with
$(J_x=\pi/4, J_y=\pi/2, J_z= \pi/4)$ acts on $|+\rangle \otimes \frac{1}{\sqrt{2}}(|00\rangle + |11\rangle)$,  the GGM of the resulting state reaches maximum \(0.5\).  However, the operator with \(
(J_x=\pi/4, J_y=11\pi/40, J_z= \pi/40)\) cannot find 
any biseparable state from \(\mathbb{S}^2\) which can create maximal GGM.


On the other hand,  
starting from  a fully separable state, \(|000\rangle\)   and \(\frac{1}{\sqrt{2}} (|0\rangle + |1\rangle) \frac{1}{\sqrt{2}}(|00\rangle + |11\rangle)\) as a biseparable input, \(U_{sp}^2\) is capable of generating maximal GGM, when \(J_i\)s satisfy certain conditions. In particular, for the fully separable state, \(J_1 - J_2 = \pi/4\) and \(J_3 + J_4 =0\) lead to maximal GME,  while the conditions in the case of the biseparable state are \(J_2 =\pi/4\) and \(J_3 - J_4 = \pi/4\).


\section{Statistical analysis of multipartite entangling power of unitary operators }
\label{sec:3qubitsresults}

Let us now move to consider more general scenarios. In particular,
for a few classes of unitary operators,  we want to assess their entangling power in terms of GGM after performing optimizations over the sets of fully separable and biseparable states. 
In contrast to our previous studies, we remove the constraint of achieving maximum GGM in this section.  
More specifically, the previous section was devoted to identify input states, both fully separable and biseparable, which are capable of creating states having maximum GGM, \(0.5\), after the action of different kinds of unitary operators. The study also revealed the interplay between the entangling power of unitaries and the entanglement present in the inputs. 
The following analysis, carried out on generic states and unitary operators, can shed some more light on this interplay. 


\subsection{Entangling capability of diagonal quantum gates}
\label{subsec:diagonal}

\begin{figure}
\includegraphics[width=1.0\linewidth]{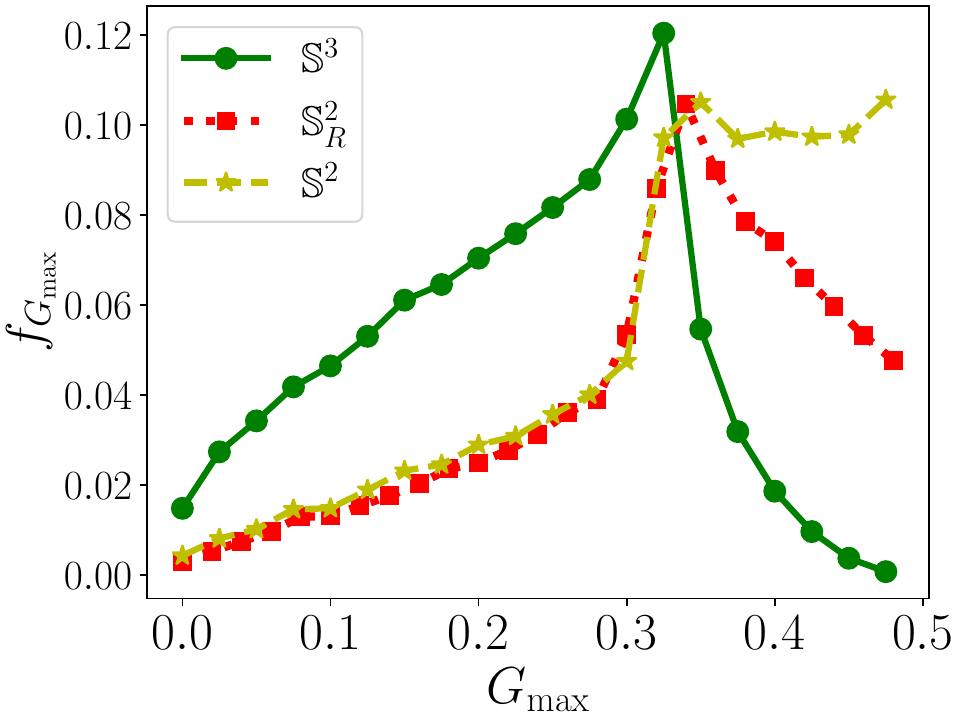}
\caption{(Color online.) Normalized frequency distribution, \(f_{G_{\max}}\) (vertical axis) \adi{as calculated via Eq. (\ref{eq:Nfreq})} against \(G_{\max}\) (horizontal axis) where the unitaries are chosen to \adi{be diagonal, \(U_D^{gen}\) by generating  \(\phi_i \in [0, 2\pi]\)  randomly} from uniform distribution. {\color{black} Stars correspond to \(f_{G_{\max}}\), obtained by optimizing over the set of fully separable states, \(\mathbb{S}^2\),  circles represent the distribution when optimizing over all biseparable states in the \(1:23\) bipartition, \(\mathbb{S}^3\) and the squares depict the same when the maximization is restricted to the set of biseparable states in \(1:23\) bipartition with only real coefficients (having vanishing \(\xi'_i)\)s), denoted as \(\mathbb{S}^2_R\). The patterns of \(f_{G_{\max}}\) do not alter if bipartitions of the biseparable states get changed and if the coefficients of the fully separable states are real. }
Both the axes are dimensionless.  }
\label{fig:diagonal} 
\end{figure}

A diagonal unitary matrix is a special kind of unitary operator having only nonvanishing diagonal elements, \(U_{D}^{gen}=\mbox{diag}(e^{i\phi_1}, \ldots, e^{i \phi_8} )\). To find \(G_{\max}\),   we randomly choose \(\phi_i\)s from uniform distribution over \([0, 2\pi]\). As we have shown in the preceding section (SubSec. \ref{subsec:entpowerdiag}), for a class of diagonal unitary operators, given by
 \(U_{D}= \mbox{diag}(1,1,1,1,1,1, 1, e^{i\phi})\),  \(G_{max}^2\) obtained from the set of biseparable states is always higher than \(G_{\max}^3\) obtained from the set of fully separable states for all values of \(\phi\) (as depicted in Fig. \ref{fig:diagonalsp}). Let us now elaborate on whether such hierarchies are maintained for randomly generated diagonal unitary operators.

 We choose a set \(\{\phi_i\}_{i=1}^{8}\) of random values of $\phi_i$s to generate a fixed  eight-dimensional diagonal unitary operator \(U_{D}^{gen}\). \adi{We repeat this process for \(5\times 10^4\) sets of random choices of \(\phi_i\)s in the diagonal unitary operators. For that operator, we compute the multipartite entangling power \(G_{\max}^i\) (i=2,3) by performing maximization over the inputs. Moreover, while performing optimizations, we consider three sets containing only (1)  fully separable states, (2) biseparable states in the
 \(1:23\)-bipartition 
 with all real coefficients, and (3) biseparable states in the \(1:23\) bipartition. We notice that the statistical behaviors of the entangling power do not qualitatively change if we maximize over the sets containing biseparable states in other bipartitions like \(3:21\) and \(2:13\). }
To analyze the situation
more carefully, we study the normalized frequency distribution of \(G_{\max}\), denoted as \(f_{G_{\max}}\). Mathematically, 
\begin{eqnarray}
    f_{G_{\max}} = \frac{\text{Number of states having}\, G_{\max}}{\text{Total number of states simulated}},
    \label{eq:Nfreq}
\end{eqnarray} 
which is depicted in Fig. \ref{fig:diagonal}. 
Examining the averages of the distributions, we find that  {\color{black}\(\overline{G}_{\max}^2 =0.33\) when the input states are restricted to the set of biseparable states with real coefficients, denoted by \(\mathbb{S}^2_R\). This value increases to \(0.35\) when optimization is performed over the set of all possible biseparable input states. In contrast, the average entanglement reduces to \(\overline{G}_{\max}^3 =0.24\)\footnote{\adi{We check the convergence of the results by simulating different numbers of unitary operators,  \(5 \times 10^3, 10^4\) and \(5\times10^4\),  with different generated random numbers. In particular, we find that the means of the distribution and the curves in Fig. \ref{fig:diagonal} do not alter with different simulated data.}} when the optimization is restricted to fully separable states. Further, we notice that in the case of fully separable inputs, if we restrict to the set of states with real coefficients,  the curve in Fig. 
\ref{fig:diagonal} and the mean of the distribution do not alter.}  These results indicate that, on an average, biseparable states are more effective inputs to create high GGM states than the fully separable ones in the case of diagonal quantum gates. 

\subsection{Specific kinds of unitary operators}
\label{subsec:specific}

\begin{figure}
\includegraphics[width=0.95\linewidth]{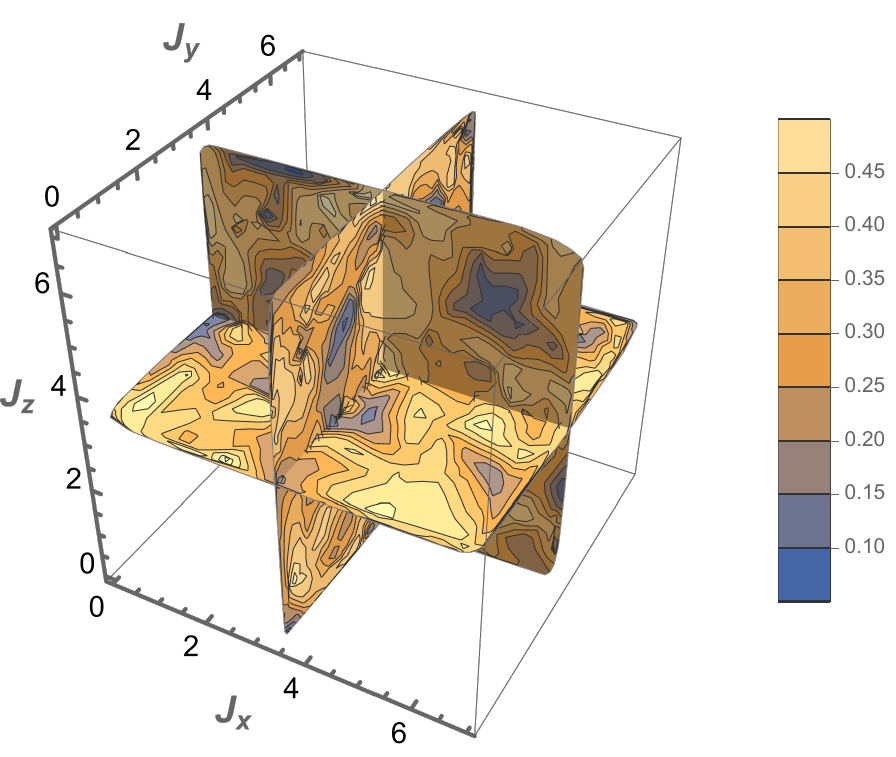}
\caption{(Color online.) 
Contour plots  in slices of multipartite entangling power, \(G_{\max}^3\) of \(U_{sp}^1\), given in Eq. (\ref{eq:spUFarook}) with respect to the parameters \(J_x\) (\(x\)-axis), \(J_y\) (\(x\)-axis) and \(J_z\) (\(z\)-axis). The optimization is performed over the set of fully separable states. 
Both the axes are dimensionless. }
\label{fig:spU1}
\end{figure}

\begin{figure}
\includegraphics[width=1.0\linewidth]{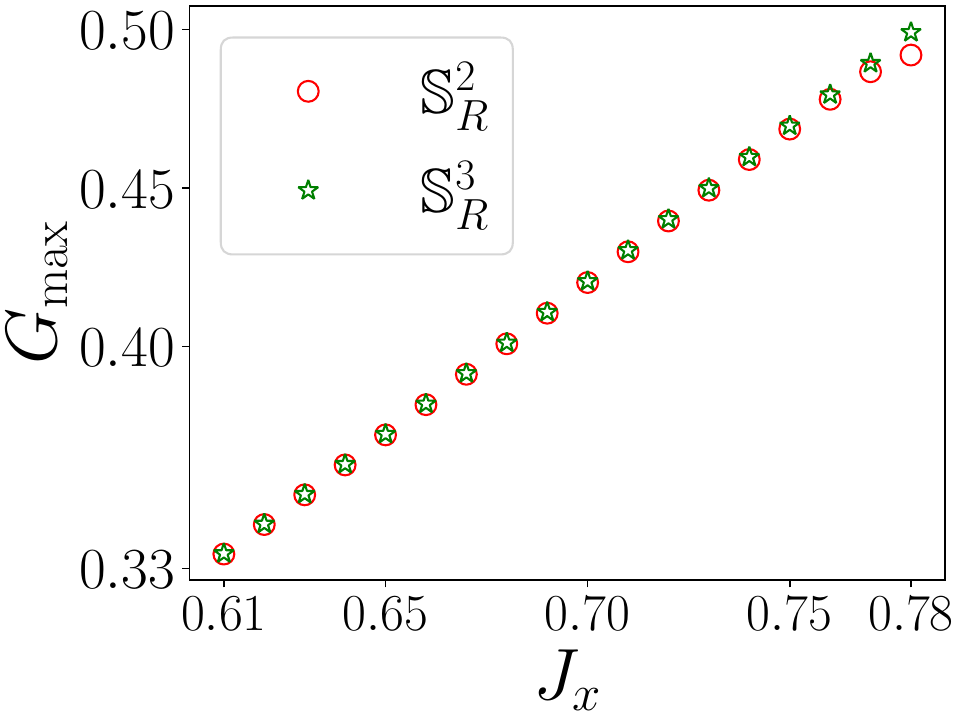}
\caption{(Color online.) {\bf Biseparable vs  fully separable states as inputs.} When a special unitary operator, given in Eq. (\ref{eq:spU}) acts on fully separable (triangles)  or biseparable states (circles), the maximum GGM, \(G_{\max}\) (ordinate) produced is plotted with respect to \(J_x\) (abscissa) for fixed \(J_y = J_z =0.1\).  This is an example of an unitary operator for which higher GGM  can be obtained from fully separable inputs compared to the biseparable ones for high values of \(J_x \sim \pi/4\) provided the coefficients of the biseparable states are chosen to be real.  Both the axes are dimensionless. }
\label{fig:spUFarookhN}
\end{figure}



We study entangling powers of two different classes of eight-dimensional unitary operators -- (1) the unitary operators \(U_{sp}^1\) in SubSec. \ref{subsec:entspecial} and (2) a certain class of unitary operators, again involving Pauli matrices.  In contrast to unitary operators described in {\color{black}SubSecs. \ref{subsec:entpowerdiag} and  \ref{subsec:entspecial},}  we illustrate some unitary operators belonging to these classes that can create higher GME states from fully separable inputs compared to biseparable ones.

Let us first determine the entangling power of \(U_{sp}^1 = \exp [i  (\sum_{i=x,y, z} J_i \sigma_i \otimes \sigma_i \otimes \sigma_z )]\), given in Eq. 
 (\ref{eq:spUFarook}). In the previous section, we have found the condition on the parameters $J_i$s of \(U_{sp}^1\) which leads to  \(G_{\max}^3 =0.5\) from a fully separable state although \(G_{\max}^2 <0.5\) for such operators, \(\tilde{U}_{sp}^1\). Hence,  fully separable states are better inputs for this class of unitary operators, \(\tilde{U}_{sp}^1\), than the biseparable ones. Let us find out \(G_{\max}^3\) for arbitrary \(U_{sp}^1\) in the   \((J_x, J_y, J_z)\)-slices. In Fig. \ref{fig:spU1}, we illustrate the contours of the  entangling power, \(G_{\max}^3(U_{sp}^1)\) by varying  \((J_x, J_y, J_z)\). It clearly manifests that these special unitaries can typically create a moderate amount of GGM from fully separable inputs, except for some small regions.

\adi{ Let us consider the other class of special eight-dimensional unitary operators,  which can 
be represented as }
\begin{eqnarray}
    U_{sp}^3 = \exp (-i \sum_{i=x,y,z} J_i \sigma_i \otimes \sigma_i\otimes \sigma_i),
    \label{eq:spU}
\end{eqnarray}
where \(J_i\)s can take any real value.  \adi{Note that the three-body interacting Hamiltonian used to construct the evolution operator has been extensively studied,  showing unique entanglement features \cite{peng_2009_prl, Shi2009} and is helpful to build quantum simulators \cite{pachos_2004,Barreiro2011}. These studies motivate us to investigate the entangling powers of the above class of unitary operators. }
By fixing the values of \(J_i\)s,  a single unitary operator is obtained for which 
\(G^i_{\max}\) have to be determined by 
maximizing  over the sets \(\mathbb{S}^i\), $i=2,3$. 

{\it Case 1 (exactly one $J_i\neq0$).} Consider \(U_{sp}^3\) for which exactly one of the \(J_i\)s is nonvanishing. In this case, for $J_i\neq 0$; $i=x,y,z$, we have  \(G_{\max}^3 = \cos^2 J_i\)   when \( (4n+1)\pi/4\leq J_i \leq (4n+3)\pi/4\) and it is \(\sin^2 J_i\) when \((4n+3)\pi/4 \leq J_i \leq (4n+5) \pi/4\), $n$ is any integer (see Appendix \ref{sec:spaunitary}). Note that it achieves maximum GGM when the nonvanishing \(J_i =n\pi/4\) (\(n=\pm 1, \pm 3, \ldots\)). 

{\color{black}When we optimize on only biseparable states, it also exhibits almost a similar amount of entangling power as \(U_{sp}^3\), although not exactly the same. We notice that the optimal biseparable input states are very close to fully separable. Suppose we consider the biseparable state of the form \(|1\rangle \otimes (\cos \theta/2 |00\rangle + \sin \theta/2 |11\rangle \). For a single nonvanishing \(J_i\), we find that it gives almost the same amount of GGM as  \(|100\rangle\) when \(\theta \) is chosen to be very small. In fact,  \(G_{\max}^2\) converges to \(G_{\max}^3\) when \(\theta \rightarrow 0\). }

{\color{black} {\it Case 2 (more than one $J_i\neq0$).} When more than one of the $J_i$s are nonvanishing, we do not have such simplified results like we had in the previous case, the picture is more involved here.} 
 {\color{black}To gain some knowledge about their entangling power, let us choose \(J_y = J_z =0.1\) while \(J_x\) varies. Like the operators \(U_{sp}^1\), these are another class of unitary operators, which can create higher GGM from real fully separable inputs in comparison to the real biseparable ones (see Fig. \ref{fig:spUFarookhN}). For example, when \(J_x=\pi/4\), from the set of fully separable states, one obtains \(G_{\max}^3 = 0.495\) while it is \(0.469\) when optimization is performed on \(\mathbb{S}^2_{R}\).}

\subsection{Randomly generated unitary operators}
\label{subsec:Haaruniform}

\begin{figure}
\includegraphics[width=1.06\linewidth]{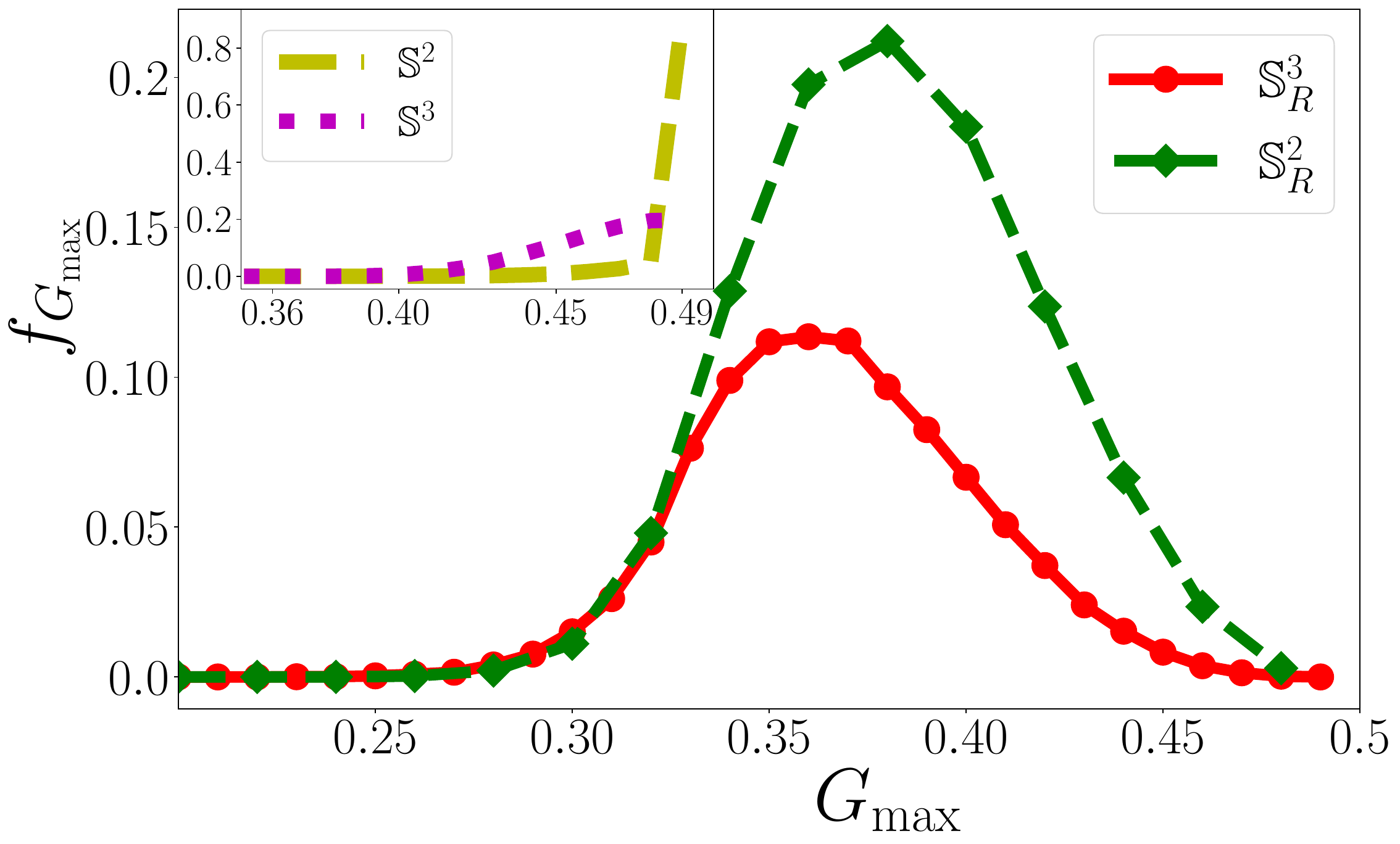}
\caption{(Color online.) \(f_{G_{\max}}\) (ordinate) against \(G_{\max}\) (abscissa) where the unitary, \(U(8)\)  is generated Haar uniformly. Diamonds  and Circles correspond to  \(f_{G_{\max}}\) obtained from the sets of real fully separable states, \(\mathbb{S}^3_R\) and real biseparable states in the \(1:23\) bipartition, \(\mathbb{S}^2_R\) respectively. In the inset, the  (magenta) dotted and (yellow) dashed curves represent \(f_{G_{\max}}\) computed with the sets of all fully separable and all biseparable states in the \(1:23\), i.e., with  \(\mathbb{S}^3\) and \(\mathbb{S}^2\) respectively.  As mentioned before, changing the set that contains biseparable states in other bipartitions does not have any qualitative effects on the results.  Both axes are dimensionless. }
\label{fig:randU} 
\end{figure}

{\color{black}We scrutinize the situation when \(10^4\) Haar uniformly generated unitary matrices \cite{KZyczkowski_1994} \(U(8)\)  act on the sets of fully separable (\(\mathbb{S}^3_R\) and \(\mathbb{S}^3\))~\footnote{\adi{The subscript \(R\)  is used to denote the coefficients of the states in the set can only be real while no subscript indicates the arbitrary quantum states, having complex coefficients.}} and biseparable input states 
(\(\mathbb{S}^2_R\) and \(\mathbb{S}^2\))}. We perform optimization over inputs for each and every choice of such a unitary operator. 
Like diagonal quantum gates, randomly generated unitary operators acting on optimal biseparable states are again capable of producing higher GME states than that from the optimal fully separable states. In this case, the normalized frequency distribution, defined in Eq. (\ref{eq:Nfreq}), depicted in  Fig. \ref{fig:randU} confirms that to generate high GGM in the resulting states, initial entanglement can be useful. Moreover, we observe that GGM created by Haar uniformly generated unitary operators is, on an average, higher than what can be obtained by the action of diagonal unitary operators. 
In this analysis, we compare the average values of the maximal multipartite entanglement generated from different classes of input states. \adi{When the input states are taken from the set of biseparable states in \(1:23\) partition with real coefficients, the average entangling power is found to be \( \overline{G}_{\max}^2 = 0.44 \) which turns out to be \( \overline{G}_{\max}^2 = 0.49 \) when the optimization is performed over the set of all biseparable states in the complex space, indicating that complex inputs are able to generate more entanglement on average under the action of the unitary operator than the states with real coefficients. Again, the results remain unaltered considering other bipartitions.}

Similarly, when we consider fully separable pure states, \adi{the average entangling power is \( \overline{G}_{\max}^3 = 0.39 \) for real inputs  and rises to \( \overline{G}_{\max}^3 = 0.46 \) for complex ones. In both situations, this  trend suggests that allowing complex coefficients in the input states enhances the unitary operator’s ability to generate genuine multipartite entanglement (cf. \cite{Sun_2025, Guo2025}).}

\begin{figure}
\includegraphics[width=1.06\linewidth]{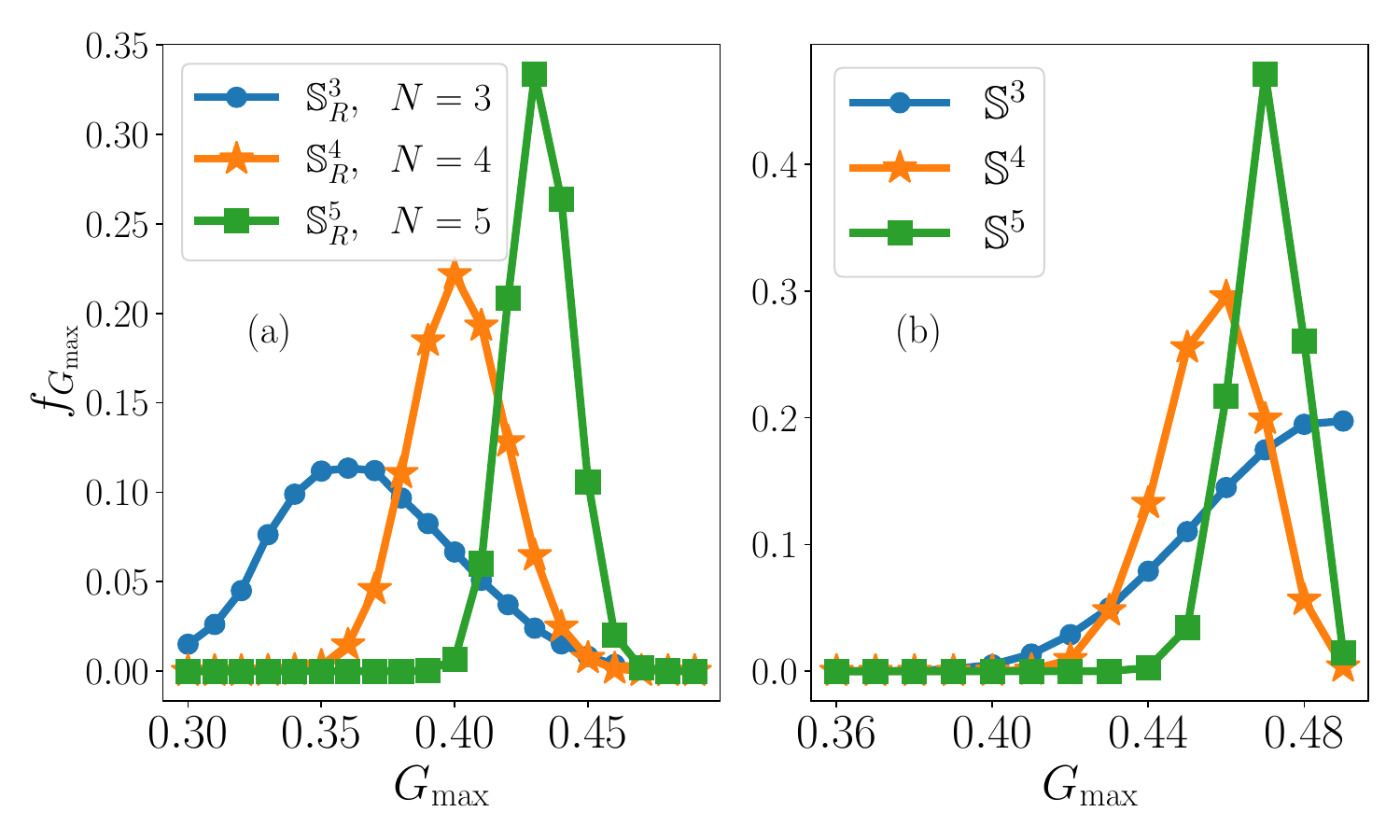}
\caption{(Color online.) Normalized frequency distribution \(f_{G_{\max}}\) (ordinate) against \(G_{\max}\) (abscissa) where the unitaries, \(U(2^N)\) with \(N=3,4,5\)  are generated Haar uniformly.   (a) Circles, stars and squares correspond to  \(f_{G_{\max}}\) obtained from the sets of fully separable states with real coefficients, \(\mathbb{S}^3_R\), \(\mathbb{S}^4_R\) and \(\mathbb{S}^5_R\) respectively. (b) The similar study is performed when the set for optimization is carried out over arbitrary fully separable states, \(\mathbb{S}^3\), \(\mathbb{S}^4\) and \(\mathbb{S}^5\). Comparing (a) and (b), we observe that the overall picture remains same in both situations, although more multipartite entanglement can be generated from arbitrary fully separable states when the dimension increases.     Both axes are dimensionless. }
\label{fig:4_5_6_full} 
\end{figure}

\adi{Going beyond eight-dimensional unitaries, we compute multipartite entangling powers for Haar uniformly simulated \(2^4\)- and \(2^5\)-dimensional unitary operators. Firstly, we investigate their normalized frequency distribution, \(f_{G_{\max}}\), when the optimization is performed over the set of fully separable states with real and complex coefficients, i.e., by choosing the sets \(\mathbb{S}^4_R\), and  \(\mathbb{S}^4\)  as well as  \(\mathbb{S}^5_R\) and \(\mathbb{S}^5\).  The patterns of \(f_{G_{\max}}\) remains same with the increase of dimension. The comparison between \(2^3\)-, \(2^4\)- and \(2^5\)-dimensional  Haar random unitaries reveal that the mean of the distribution increases, although standard deviation decreases as the dimension grows (see Fig. \ref{fig:4_5_6_full}). Specifically, for four quibits, \(\overline{G}^4_{\max } = 0.40\) with \(\mathbb{S}^4_R\) and  \(\overline{G}^4_{\max } =0.46 \) after optimizing over all possible fully separable states. In the case of five qubits, we find \(\overline{G}^5_{\max } =0.43 \) and 
\(\overline{G}^5_{\max } =0.47 \) for the maximization set \(\mathbb{S}^5_R\) and \(\mathbb{S}^5\) respectively. }

\section{Conclusion}
\label{sec:conclu}

Entanglement generated at various stages of quantum information processing tasks plays a central role in enabling quantum advantages.
To implement a quantum algorithm, a quantum circuit must be designed, consisting of a sequence of quantum gates represented by unitary operators. The entangling power of a fixed quantum gate can be defined as the average or maximum amount of entanglement it produces when acting on a set of product states.
For unitary operators acting on multipartite separable states, genuinely multipartite entangled (GME) states can, in general, be created, thereby revealing their entangling capability. However, in the multipartite setting, there exist different classes of separable states, including fully separable states, which contain no entanglement in any bipartition, and separable states having entanglement in at least one bipartition.

 In literature, the entangling power of a unitary operator in a multipartite system was determined by its entanglement generation capability in its reduced subsystems or in all of its bipartitions \cite{Scott04, Linowski_2020}.  Going beyond this conventional approach, we quantified multipartite entangling power of a given  unitary operator via its potential to create  genuine multipartite entangled outputs after the action on   multipartite separable states.
 We proposed that the way the operators act on various classes of separable states can be utilized to introduce the concept of hierarchies among those operators. To make the hierarchies meaningful, we suggested that the sets of states on which the maximization is performed should ideally be disjoint. {\color{black} We also proved features of this multipartite entangling power including its invariance  under the local unitary transformations.} 

In order to generate genuine multipartite entanglement, we find that the input entanglement present in some of the bipartitions of the input states is not always helpful. Specifically, we identified several classes of unitary operators and their corresponding \(k\)-separable inputs which lead to maximum genuine multipartite entangled states as outputs.  
  We discovered that in three-qubits, there are unitary operators such as diagonal and Haar uniformly generated unitary operators that can  result in the generation of higher genuine multipartite entanglement, on an average, from biseparable states than that from the fully separable ones. However, the opposite also holds on a substantial number of occasions. \adi{These findings demonstrate that the relationship between input-state separability and output GME is far from straightforward. Furthermore, we found that the entangling power obtained by optimizing over states with only real coefficients is consistently lower than that achieved using arbitrary separable states with complex coefficients.
Our studies introduce a novel classification framework for multipartite unitary operators in terms of their capability to create genuine multipartite entanglement and on the classes of separable inputs that maximize this power.}

\acknowledgements
 The authors acknowledge the support from the Interdisciplinary Cyber Physical Systems (ICPS) program of the Department of Science and Technology (DST), India, Grant No.: DST/ICPS/QuST/Theme- 1/2019/23 and the use of cluster computing facility at the Harish-Chandra Research Institute.   M.S., S.M and A.S.D. acknowledge the support from the project entitled "Technology Vertical - Quantum Communication'' under the National Quantum Mission of the Department of Science and Technology (DST)  ( Sanction Order No. DST/QTC/NQM/QComm/2024/2 (G)).

\appendix

\section{Entangling power of diagonal unitary for biseparable inputs }
\label{sec:bi_entagling_power}

 We consider a biseparable input state, which is separable between the first qubit and the joint state of the remaining two qubits. Hence, the initial state is written as
\begin{eqnarray}
    \ket{\psi^{2\text{-sep}}}_3 = \ket{\psi_1}\otimes\ket{\psi_{23}},
\end{eqnarray}
where
\begin{eqnarray}
   \nonumber \ket{\psi_1}& = &\cos\frac{\theta_1}{2}\ket{0} + e^{i\xi_1}\sin\frac{\theta_1}{2}\ket{1},\\
   \nonumber \text{and}\, \, \ket{\psi_{23}}&=& \cos{\frac{\theta_2}{2}}\ket{00} + e^{i\xi_2}\sin{\frac{\theta_2}{2}}\cos{\frac{\theta_3}{2}}\ket{01}\\ \nonumber &+& e^{i\xi_3}\sin{\frac{\theta_2}{2}}\sin{\frac{\theta_3}{2}}\cos{\frac{\theta_4}{2}}\ket{10}\\\nonumber &+& e^{i\xi_4}\sin{\frac{\theta_2}{2}}\sin{\frac{\theta_3}{2}}\sin{\frac{\theta_4}{2}}\ket{11}.
\end{eqnarray}
Upon application of the global unitary \(U_D\), the output state becomes 
\begin{eqnarray}
    \ket{\psi_{out}}=U_D \ket{\psi^{2\text{-sep}}}_3,
\end{eqnarray}
Our numerical analysis confirms that the GGM is fully determined by the reduced density matrix \(\rho_1\) of the first qubit. The explicit form reads as
\begin{eqnarray}
    \rho_1 = \begin{pmatrix}
p && q \\
q^* && r
    \end{pmatrix},
\end{eqnarray}
with the matrix elements given by
\(p=\cos^2 \frac{\theta_1}{2},\quad
 q=l_1  e^{-i \xi_1} \cos\frac{\theta_1}{2} \sin\frac{\theta_1}{2} ,\quad
 r= \sin^2 \frac{\theta_1}{2}
\). Here \(l_1=\cos^2 \frac{\theta_2}{2} + l_2
\sin^2 \frac{\theta_2}{2}, \quad l_2=\cos^2 \frac{\theta_3}{2} + l_3
\sin^2 \frac{\theta_3}{2}, \quad \text{and}\, l_3=\cos^2 \frac{\theta_4}{2} + e^{-i\phi} 
\sin^2 \frac{\theta_4}{2}\).
Its eigenvalues are
\begin{eqnarray}
    \lambda^\pm = \frac{1}{2}\pm \frac{\sqrt{1-4(p r - |q|^2)}}{2}.
 \label{eigenvalue_1st}
\end{eqnarray}
From  Eq. (\ref{eigenvalue_1st}), it is clear that the eigenvalues and, therefore, the GGM do not depend on the local phase \(\xi_i\), since the eigenvalues depend only on \(|q|\). Hence, the GGM of this output state is \(\lambda^-\). If we calculate the maximum value of the GGM of this unitary, we seek to maximize \(\lambda^-\), which occurs when the quantity \((pr-|q|^2)\) is maximized. The conditions for maximization are
\begin{eqnarray}
   \frac{d(pr-|q|^2)}{d\theta_i} =0,\quad
  \frac{d^2(pr-|q|^2)}{d^2\theta_i} <0.
  \label{condition_max_theta_i}
\end{eqnarray}
For \(i=1\), the first and second derivatives are 
\begin{widetext}
\begin{eqnarray}
 \nonumber   \frac{d(pr-|q|^2)}{d\theta_1}= \frac{1}{8} \sin 2\theta_1 \, \sin^2\frac{\theta_2}{2} \, \sin^2\frac{\theta_3}{2} \, \sin^2\frac{\theta_4}{2} \left(7 + \cos \theta_3 + \cos \theta_4 - \cos \theta_3 \cos \theta_4 + 4 \cos \theta_2 \sin^2\frac{\theta_3}{2} \sin^2\frac{\theta_4}{2}\right) \sin^2\frac{\phi}{2}, \\
 \nonumber \frac{d^2(pr-|q|^2)}{d^2\theta_1} = \frac{1}{4} \cos 2\theta_1 \, \sin^2\frac{\theta_2}{2} \, \sin^2\frac{\theta_3}{2} \, \sin^2\frac{\theta_4}{2} \left(7 + \cos \theta_3 + \cos \theta_4 - \cos \theta_3 \cos \theta_4 + 4 \cos \theta_2 \sin^2\frac{\theta_3}{2} \sin^2\frac{\theta_4}{2}\right) \sin^2\frac{\phi}{2}.\\
 \label{condition_theta_1}
\end{eqnarray}
\end{widetext}
From Eqs. (\ref{condition_max_theta_i}) and (\ref{condition_theta_1}), the quantity \((pr-|q|^2)\) is maximum when  \(\theta_1=\frac{\pi}{2}\)  in the range \(0\leqslant \theta_    1\leqslant \pi\). For the remaining angles \(\theta_i\) with \(i=\{2,3,4\}\), the first derivative is
\begin{eqnarray}
    \frac{d(pr-|q|^2)}{d\theta_i} &=& \frac{1}{8} \sin^2 \theta_1  \sin \theta_i  \sin^2 \frac{\theta_{i+1}}{2}  \sin^2 \frac{\theta_{i+2}}{2} \nonumber\\&&( 4 - 8 \sin^2 \frac{\theta_2}{2}  \sin^2 \frac{\theta_3}{2}  \sin^2 \frac{\theta_4}{2} ) \sin^2 \frac{\phi}{2},\nonumber\\
\end{eqnarray}
where the indices are cyclic, if  \(i=4\), then \(i+1=2\) and \(i+2=3\). The maximum occurs when the term inside the parentheses vanishes, i.e.,  
\begin{eqnarray}
     4 - 8 \sin^2 \frac{\theta_2}{2}  \sin^2 \frac{\theta_3}{2}  \sin^2 \frac{\theta_4}{2}=0.
     \label{theta_2-theta_3_theta_4}
\end{eqnarray}
 Under this condition (Eq. (\ref{theta_2-theta_3_theta_4})), the second derivative in Eq. (\ref{condition_max_theta_i})  becomes
 \begin{eqnarray}
  \nonumber   \frac{d^2(pr-|q|^2)}{d^2\theta_i} &=& -\frac{1}{2} \sin^2 \theta_1 \sin^2\theta_i\sin^2\frac{\phi}{2}\sin^2\frac{\theta_{i+1}}{2}\sin^2\frac{\theta_{i+2}}{2}\\&<&0, \quad  i=\{2,3,4\}\\ \nonumber
 \end{eqnarray}
The optimal input biseparable state of this unitary operator satisfies \(\theta_1=\frac{\pi}{2}\) and Eq. (\ref{theta_2-theta_3_theta_4}). In our case, we choose \(\theta_2 =\frac{\pi}{2},\,\theta_3=\pi,\,\theta_4=\pi\). This gives the optimal biseparable input state as
\begin{eqnarray}
    \ket{\psi^{2-sep}_{opt}}_3 = \frac{1}{\sqrt{2}}(\ket{0}+\ket{1})\frac{1}{\sqrt{2}}(\ket{00}+\ket{11}).
\end{eqnarray}
After applying the unitary \(U_\phi\) on this optimal state \(\ket{\psi^{2-sep}_{opt}}_3\), the output state is written as 
\begin{eqnarray}
    \ket{\psi^{opt}_{out}} = \frac{1}{2}(\ket{000}+\ket{011}+\ket{100}+e^{i\phi}\ket{111}).
\end{eqnarray}
For this output state, the reduced density matrix of the first qubit is written as 
\begin{eqnarray}
    \rho_1 = \begin{pmatrix}
\displaystyle \frac{1}{2} 
&
\displaystyle \frac{1}{4}(1 +  e^{-i\phi}) \\[6pt]
\displaystyle \frac{1}{4}(1 + e^{i\phi})
&
\displaystyle \frac{1}{2}
\end{pmatrix},
\end{eqnarray}
and the eigenvalue of this density matrix are \(\lambda_{1,2}=[\sin^2\frac{\phi}{4},\cos^2\frac{\phi}{4}]\). In this case, \(\rho_2 =\rho_3 = \mathbb{I}/2\) and hence GGM turns out to be \(\max[\sin^2\frac{\phi}{4},\cos^2\frac{\phi}{4}]\). 


\section{Entangling power of diagonal unitary when the input states are fully separable }
\label{sec:full_entagling_power}

To calculate the entangling power \({G}^3_{\text{max}}(U_{D})=\underset{|\psi^{3\text{-sep}} \rangle_{3} \in \mathbb{S}_3}{\max} G(U_{D} |\psi^{3\text{-sep}} \rangle_{3})\), we have to calculate   the reduced density  matrix \(\rho_i\) \((i=1,2,3\)) for each party of the given state in Eq. (\ref{equ:full_out_diag_one}).  
They can take the form as 
\begin{equation}
    \rho_i = \begin{pmatrix}
    a_i & b_i\\
    b^*_i & c_i
    \end{pmatrix},
\end{equation}
where 
\begin{eqnarray}
 \nonumber   a_i &=&{\cos^2\frac{\theta_i}{2}} ,\\
  \nonumber  b_i &= &\cos\frac{\theta_i}{2} \sin\frac{\theta_i}{2} \left( \sum_{\substack{j=1 \\ j \neq i}}^{3-\min(i,2)}
 \cos^2\frac{\theta_j}{2} + r_1\sin^2\frac{\theta_j}{2}  \right),
\\
  \text{and} \,  c_i&=& {\sin^2\frac{\theta_i}{2}},
\end{eqnarray}
where \(r_1=\left( \overset{3}{\underset{\substack{k=1 \\ k \neq j,i}}{\sum}} \cos^2\frac{\theta_k}{2}+ e^{-i\phi} \sin^2\frac{\theta_k}{2} \right)\).
The eigenvalues of the reduced density matrix \(\rho_i\) are then given by
\begin{eqnarray}
    \lambda^\pm_i = \frac{1}{2}\pm \frac{\sqrt{1-4(a_i c_i - |b_i|^2)}}{2}.
    \label{eigenvalue}
\end{eqnarray}
 Therefore, the GGM of the output state is $G=(1-\max\{\lambda^+_1,\lambda^+_2,\lambda^+_3\})$. To find the maximum GGM that the unitary \(U\) can generate, we optimize over all possible input product states, i.e., 
 \begin{eqnarray}
  \nonumber  G^3_{\max}&=& \underset{\theta_i}{\max} ~~G\\
 \nonumber &=& \underset{\theta_i}{\max}(1-\max\{\lambda^+_1,\lambda^+_2,\lambda^+_3\})\\
 \nonumber &=& 1- \underset{\theta_i}{\min}(\max\{\lambda^+_1,\lambda^+_2,\lambda^+_3\})\\
 \nonumber &=& 1-\max\{\underset{\theta_i}{\min}(\lambda^+_1,\lambda^+_2,\lambda^+_3)\}\\
 &=& 1- \max\{\underset{\theta_i}{\min}~\lambda^+_1,\underset{\theta_i}{\min}~\lambda^+_2,\underset{\theta_i}{\min}~\lambda^+_3\},
 \end{eqnarray} 
which implies that maximizing the GGM is equivalent to maximizing the quantity $(a_i c_i - |b_i|^2)$ for each qubit.
The condition for this maximization can be written as
\begin{eqnarray}
  \nonumber  \frac{d(a_ic_i-|b_i|^2)}{d\theta_j} &=&0, \\  \frac{d^2(a_ic_i-|b_i|^2)}{d\theta_j^2} &<&0,\quad \forall i,j\in\{1,2,3\}.
  \label{eq:der_3qubit}
\end{eqnarray}
There are many possibilities of Eq.~(\ref{eq:der_3qubit}) but the suitable condition for the maximization is given by
\begin{eqnarray}
    \nonumber  \frac{d(a_ic_i-|b_i|^2)}{d\theta_{i+1}} &=&0, \forall i\in\{1,2,3\}\\  \frac{d^2(a_ic_i-|b_i|^2)}{d\theta_{i+1}^2} &<&0.
  \label{eq:der_3qubit_suitable}
\end{eqnarray}

From Eq.~(\ref{eq:der_3qubit_suitable}), one can arrive at  conditions,
\begin{eqnarray}
\nonumber 1+\cos{\theta_{i+1}}+\cos{\theta_{i+2}}-\cos{\theta_{i+1}}\cos{\theta_{i+2}} = 0\\ \forall i \in\{1,2,3\},
     \label{eq:con_der1}
\end{eqnarray}
and
\begin{eqnarray}
  \nonumber  -\frac{1}{2} \sin^2\theta_i \sin^2\theta_{i+1} \sin^4\frac{\theta_{i+2}}{2} \sin^2\frac{\phi}{2}<0,
 \\\nonumber \forall i\in \{1,2,3\}.
\end{eqnarray}
Note that from Eq. (\ref{eq:con_der1}),
we reach the condition  \(\cos\theta_1=\cos\theta_2=\cos\theta_3\), i.e., \(\theta_1=\theta_2=\theta_3\) and the eigenvalues take the form as 
\begin{eqnarray}
\lambda_\pm&=&\frac{1}{2}\pm \frac{1}{32}\sqrt{A}.
    \label{eq:eigval_3sep}
\end{eqnarray}
 Here \(A=218 + 16 \cos\theta + 49 \cos2 \theta - 24 \cos3 \theta - 10 \cos 4 \theta + 8 \cos 5 \theta - \cos 6 \theta   - 1024 \cos^4\frac{\theta}{2}\left(-3 + \cos\theta\right) \cos\phi \sin^6\frac{\theta}{2}.\) Thus, the entangling power, \({G}^3_{\max}(U_{D})=\underset{\theta}{\max}(\frac{1}{2}-\frac{1}{32}\sqrt{A})\). Through numerical analysis, we find that when \(\phi=\pi\),  the entangling power, \({G}^3_{\max}(U_{D}(\phi=\pi))=0.34\) upon optimizing over \(\theta\).

\newpage
\section{Entangling power of spacial unitary operator for fully separable inputs}
\label{sec:spaunitary}

We begin by considering a general  fully separable input state of three qubits, parametrized as
\begin{eqnarray}
    \ket{\psi^{3-\text{sep}}}_{3} = \overset{3}{\underset{i=1}{\bigotimes}}\cos\frac{\theta_i}{2}\ket{0} + e^{i\xi_i}\sin\frac{\theta_i}{2}\ket{1},
\end{eqnarray}
where \(\theta_i\in[0,\pi]\) and \(\xi_i\in[0,2\pi]\) are the polar and azimuthal angles on the Bloch sphere for qubit \(i\). The system then undergoes a global unitary evolution governed by
\begin{eqnarray}
    U=\text{exp}(iJ_x\sigma_x\otimes \sigma_x\otimes\sigma_x).
\end{eqnarray}
To compute the GGM of the output state, we evaluate the reduced density matrices \(\rho_i\) of each individual qubit after the action of \(U\). These take the general form
\begin{eqnarray}
    \rho_i = \begin{pmatrix}
p_i && q_i \\
q^*_i && r_i
    \end{pmatrix},
\end{eqnarray}
where
\begin{eqnarray}
\nonumber p_i &=& \frac{1}{4} ( 
2 + \cos( 2J_x - \theta_i )+ \cos( 2J_x + \theta_i ) \\\nonumber
 &-& 2 \cos \xi_{i+1} \cos \xi_{i+2} \sin 2J_x \sin \theta_1 \sin \theta_2 \sin \theta_3 \sin \xi_i
),
\\
\nonumber q_i &=&  \frac{1}{2} \left( 
\cos \xi_i \sin \theta_i - 
i l_4
\right),\\
\text{and}\, \, r_i&=&1-p_i.
\end{eqnarray}
Here \(l_4=( 
\cos \theta_i \cos \xi_{i+1} \cos \xi_{i+2} \sin 2J_x \sin \theta_{i+1} \sin \theta_{i+2} 
+ \cos 2J_x \sin \theta_i \sin \xi_i )\).
The eigenvalues of the reduced density matrix \(\rho_i\) are then given by
\begin{eqnarray}
    \lambda^\pm_i = \frac{1}{2}\pm \frac{\sqrt{1-4(p_i r_i - |q_i|^2)}}{2}.
    \label{eigenvalue}
\end{eqnarray}
 Hence, the GGM of the output pure state is given by ${G}=(1-\max\{\lambda^+_1,\lambda^+_2,\lambda^+_3\})$. The quantity depends on the state parameters and hence we have to perform optimization over \(\theta_i\)s and \(\xi_i\)s.  Specifically, 
 \begin{eqnarray}
  \nonumber  G^3_{max}&=& \underset{\xi_i,\theta_i}{\max} ~~ G\\
 \nonumber &=& \underset{\xi_i,\theta_i}{\max}(1-\max\{\lambda^+_1,\lambda^+_2,\lambda^+_3\})\\
 \nonumber &=& 1- \underset{\xi_i,\theta_i}{\min}(\max\{\lambda^+_1,\lambda^+_2,\lambda^+_3\})\\
 \nonumber &=& 1-\max\{\underset{\xi_i,\theta_i}{\min}(\lambda^+_1,\lambda^+_2,\lambda^+_3)\}\\
 &=& 1- \max\{\underset{\xi_i,\theta_i}{\min}~\lambda^+_1,\underset{\xi_i,\theta_i}{\min}~\lambda^+_2,\underset{\xi_i,\theta_i}{\min}~\lambda^+_3\}.
 \end{eqnarray} 
Again, by maximizing $(p_i r_i - |q_i|^2)$, we have to ensure that the following equations are satisfied: 
\begin{eqnarray}
 \nonumber   \frac{d(p_i r_i - |q_i|^2)}{d\xi_j}=0, \quad
    \frac{d^2(p_i r_i - |q_i|^2)}{d^2\xi_j}<0, 
    \label{condition_opt}
\end{eqnarray}
 Similarly, we find out that they satisfy
\begin{eqnarray}
    \frac{d(p_i r_i - |q_i|^2)}{d\xi_{i}}=0,\quad \frac{d^2(p_i r_i - |q_i|^2)}{d^2\xi_{i}}<0.
    \label{condition_effect}
\end{eqnarray}
Finally, we have
\begin{widetext}
\begin{eqnarray}
\nonumber \frac{d(p_i r_i - |q_i|^2)}{d\xi_i} &=&-\frac{1}{8} \sin 2\xi_i \sin^2 \theta_i \left[
-1 + \cos^4 J_x + \sin^4 J_x + 2 \cos^2 J_x \sin^2 J_x \left(
-3 + 4 \prod_{\substack{j=1 \\ j \neq i}}^{3} \cos^2 \xi_j \sin^2 \theta_j
\right)
\right],  \\
 \frac{d^2(p_i r_i - |q_i|^2)}{d^2\xi_{i}} &=& -
\frac{1}{4} \cos 2\xi_i \sin^2 \theta_i \left[
-1 + \cos^4 J_x + \sin^4 J_x + 2 \cos^2 J_x \sin^2 J_x \left(
-3 + 4  \prod_{\substack{j=1 \\ j \neq i}}^{3} \cos^2 \xi_j \sin^2 \theta_j
\right)
\right].\nonumber \\
\label{derivative}
\end{eqnarray}
\end{widetext}
Again, the optimal values turn out to be $\xi_i=0$ or \(\pi\) and we have
\begin{eqnarray}
 \nonumber   {\lambda^R}^\pm_i &=& \frac{1}{8}\bigg(4\pm\sqrt{\alpha_i}\bigg),
 \end{eqnarray}
where \(\alpha_i = 13 - 3 \cos 2\theta_i + 
2 \cos^2 \theta_i (
    -\cos 2\theta_{i+2} 
    - 2 \cos 2\theta_{i+1} \sin^2 \theta_{i+2} 
+ \cos 4J_x ( 
        3 + \cos 2\theta_{i+2} 
        + 2 \cos 2\theta_{i+1} \sin^2 \theta_{i+2} 
    )
)\).
Hence, we now optimize the remaining angles \(\theta_i\) to find the maximum GGM as
\begin{eqnarray}
    G^3_{\max} = 1- \max\{\underset{\theta_i}{\min}~{\lambda^R}^+_1,\underset{\theta_i}{\min}~{\lambda^R}^+_2,\underset{\theta_i}{\min}~{\lambda^R}^+_3\}.
\end{eqnarray}
The optimal conditions are obtained by solving
\begin{eqnarray}
     \frac{d\alpha_i}{d\theta_{i}}=0,\quad \frac{d^2\alpha_i}{d^2\theta_{i}}>0,   \label{real_equation}
\end{eqnarray}
which leads to
\begin{eqnarray}
 \nonumber   \frac{d\alpha_i}{d\theta_{i}} &=& 4 l_5   \sin^2 2J_x  \sin 2\theta_i ,\\
\nonumber \frac{d^2\alpha_i}{d^2\theta_{i}} &=& 8 l_5  \sin^2 2J_x  \cos 2\theta_i .\\
\label{theta_condition}
\end{eqnarray}
Here \(l_5=3 + \cos 2\theta_{i+2} + 2 \cos 2\theta_{i+1}  \sin^2 \theta_{i+2}\).
From Eqs. (\ref{real_equation}) and  (\ref{theta_condition}), we see that \(\theta_i=0,\pi\,~~\forall i\in\{1,2,3\}\). This leads to the conclusion that the optimal input states that maximize the entanglement generated by the unitary \(U\) are either \(\ket{000}\) or \(\ket{111}\). The reduced density matrix of each qubit after the action of this unitary on this optimal input state is given by 
\begin{eqnarray}
  \rho^{opt}_i=  \begin{pmatrix}
      \cos^2 J_x && 0\\ 0 && \sin^2 J_x
    \end{pmatrix}.
\end{eqnarray}
Similar calculations occur for $J_y$ and $J_z$.

\bibliography{ref}
\end{document}